\documentclass[showpacs,onecolumn,preprintnumbers,amsmath,amsfonts,amssymb,floatfix,aps,superscriptaddress]{revtex4}

\usepackage{physymb}
\usepackage{graphicx}
\usepackage{epsfig}
\usepackage[hang,nooneline]{subfigure}
\usepackage[normalem]{ulem}
\usepackage{color}
\usepackage{hyperref}

\newcounter{fig}

\newcommand{\vect}[1]{\mathbf{#1}}

\begin{document}

\title{$SO(2)$-induced breathing patterns in 
multi-component Bose-Einstein condensates
}
\author{E. G. Charalampidis}
\email[Email: ]{charalamp@math.umass.edu}
\affiliation{Department of Mathematics and Statistics, University of Massachusetts
Amherst, Amherst, MA 01003-4515, USA}
\author{Wenlong Wang}
\email[Email: ]{wenlongwang@physics.tamu.edu}
\affiliation{Department of Physics and Astronomy, Texas A\textsl{\&}M University,
College Station, TX 77843-4242, USA}
\author{P. G. Kevrekidis}
\email[Email: ]{kevrekid@math.umass.eduu}
\affiliation{Department of Mathematics and Statistics, University of Massachusetts
Amherst, Amherst, MA 01003-4515, USA}
\author{D. J. Frantzeskakis}
\email[Email: ]{dfrantz@phys.uoa.gr}
\affiliation{Department of Physics, National and Kapodistrian University of Athens, Panepistimiopolis, Zografos,
Athens 15784, Greece}
\author{J. Cuevas-Maraver}
\email[Email: ]{jcuevas@us.es}
\affiliation{Grupo de F\'{i}sica No Lineal, Departamento de F\'{i}sica Aplicada I,
Universidad de Sevilla. Escuela Polit\'{e}cnica Superior, C/ Virgen de \'{A}frica, 7, 41011-Sevilla, Spain\\
Instituto de Matem\'{a}ticas de la Universidad de Sevilla (IMUS). Edificio Celestino Mutis. Avda. Reina Mercedes s/n, 41012-Sevilla, Spain}

\date{\today} 

\begin{abstract}
In this work, we employ the $SO(2)$-rotations of a two-component, 
one-, two- and three-dimensional nonlinear Schr\"{o}dinger system 
at and near the Manakov limit, to construct vector solitons and 
vortex structures. This way, stable stationary dark-bright solitons
and their higher-dimensional siblings are transformed into robust 
oscillatory dark-dark solitons (and generalizations thereof), with
and without a harmonic confinement. By analogy to the one-dimensional
case, vector higher-dimensional structures take the form of vortex-vortex
states in two dimensions and, e.g., 
vortex ring-vortex ring ones in three dimensions.
We consider the effects of unequal (self- and cross-) interaction
strengths, where the $SO(2)$ symmetry is only approximately satisfied, showing
the dark-dark soliton oscillation is generally robust. Similar features
are found in higher dimensions too, although our case examples suggest 
that phenomena such as phase separation may contribute to the
associated dynamics. These results, in connection with 
the experimental realization of one-dimensional variants of such states 
in optics and Bose-Einstein condensates (BECs), suggest the potential 
observation of the higher-dimensional bound states proposed herein.
\end{abstract}

\maketitle

\section{Introduction}

One of the most paradigmatic models of multi-component
system dynamics within integrable nonlinear systems
and wave phenomena is the so-called Manakov model~\cite{manakov,intman1}.
This is a vector variant of the famous nonlinear Schr{\"o}dinger (NLS)
equation~\cite{sulem,ablowitz,siambook}, featuring equal
(nonlinear) interactions within a certain component and
across the different ones. Vector solitons of this model have attracted
considerable interest, both in the case of focusing~\cite{ablowitz} and 
defocusing~\cite{siambook,rip} nonlinearities. 

In the present setting, the case that will be of interest is that of the  
defocusing nonlinearity, as referred to in nonlinear optics; on the other
hand, in the context of atomic Bose-Einstein condensates (BECs)~\cite{stringari},
this case corresponds to repulsive interatomic interactions, and is thus 
referred to as repulsive nonlinearity. In the original, one-dimensional 
(1D) Manakov system a particularly intriguing structure that is supported
is the  so-called dark-bright (DB) soliton. Here, the bright soliton component,
which would not otherwise exist in the defocusing setting, arises because 
of an effective potential well created by the dark soliton through the 
inter-component interaction. In that light, DB solitons can be considered
as ``symbiotic'' structures. The extensive study of such states~\cite{christo,vdbysk1,vddyuri,ralak,dbysk2,shepkiv,parkshin},
has stemmed in good measure from their potential applications in optics, 
where dark solitons were proposed to effectively act as adjustable waveguides
for weak signals~\cite{kivshar}. In this field, the theoretical/analytical
developments were (already a couple of decades ago) supplemented by experimental
work in photorefractive media pioneering the observation of DB 
structures~\cite{seg1,seg2}.

Our work is largely inspired by the setting of atomic condensates,
where such structures have been explored in multiple recent 
experiments. The latter, were chiefly focusing on the dynamics 
of pseudo-spinor (two-component) atomic gases, featuring two hyperfine
states of the same atom species, such as $^{87}$Rb. Here, the early 
theoretical prediction of DB solitons \cite{buschanglin} was -- after
a considerable hiatus -- followed by the experimental realization by 
the Hamburg group~\cite{hamburg}. This, in turn, led to numerous further
directions of explorations, many of which were pursued at Pullman~\cite{pe1,pe2,pe3,pe4,pe5,azu}. 
In particular, in Ref.~\cite{hamburg}, robust DB solitons were created
by a phase-imprinting technique and their robust oscillations were probed
in a quasi-1D parabolic trap. On the other hand, in subsequent 
experiments, different types of structures, including DB and dark-dark 
(DD) solitons~\cite{pe1,pe2,pe3,pe4,pe5}, emerged spontaneously via 
instabilities in counterflow dynamical scenarios.

In the BEC setting, one of the significant advantages
of the spinor gases is that, naturally, the coefficients
of inter- and intra-component interactions are very close
to being identical; in fact, the differences are not more than a few
percent, which is important, e.g., in phase separation~\cite{siambook,stringari,emergent}. Hence,
the model is naturally proximal (in as far as its nonlinearity coefficients 
are concerned) to the Manakov one. Remarkably, the Manakov
case bears an additional symmetry under rotations, i.e., the
model is invariant under the action of the $SU(2)$ Lie group.
This invariance has been employed in order to generate 
unitary (in fact, chiefly orthogonal) rotations of states, such as the 
DB solitons. The resulting waveforms, produced even experimentally~\cite{pe4,pe5}, are 
a particular form of DD solitons. Depending on the frequency (chemical
potential, as we will refer to it below), the resulting evolution
of 
the components can be intrinsically oscillatory i.e., breathing
in their atomic density. While the transformation is exact
only in the Manakov case, weak deviations from this integrable
limit relevant to the atomic species appear to maintain such DD states
as sufficiently robust nonlinear excitations in order for them
to be experimentally observable.

In the present work, we extend the consideration of such states 
to higher dimensions. In particular, in section II, we revisit 
the mathematical framework of the $SU(2)$ and more specifically 
the $SO(2)$ group generator that produces the relevant invariance.
For completeness, and also to make connections with earlier work, 
in section III.A we start with the 1D case by briefly discussing 
the DD soliton, stemming from the rotation of the DB one. We then
examine the two-dimensional (2D) case, by considering 
vortex-bright solitons. The latter, involve a vortex in one component 
trapping a bright soliton in the other component~\cite{kodyprl,pola}. 
These structures, are also known as ``filled core vortices'' (that were
experimentally observed in Ref.~\cite{anders}), ``half-quantum vortices''%
~\cite{tsubota,adhikari} or ``baby Skyrmions''~\cite{cooper}, and their
stability and dynamics, have been studied, respectively, in Refs.~\cite{skryabin,kodyprl}
and Refs.~\cite{kodyprl,tsubota}. Rotating these in section III.B, we 
obtain (by analogy to the DD states) vortex-vortex structures with their
constituent vortices 
rotating around one another. Finally, we turn to the three-%
dimensional (3D) setting, and examine the cases of vortex lines and vortex
rings, which are the prototypical excitations therein~\cite{siambook,komineas}.
Once again, our starting point is the vortex-line, or vortex-ring in the 
first component, that traps a (line or ring, respectively) bright soliton
in the second component. The rotation of such a stationary state allows 
us to capture a vector vortex-ring state with its own intrinsic vibrating
dynamics, as we will illustrate in what follows in section III.C. Finally, 
in section IV, we summarize our findings and present a number of possibilities
for future study.

\section{The model and analytical/computational setup}

We start by presenting our model, as well as the analytical 
and computational setup. We consider the coupled defocusing
NLS system written in dimensionless form~\cite{siambook} as
\begin{subequations}
\begin{eqnarray}
i\partial _{t}{\Phi _{-}} &=&-\frac{1}{2}\nabla^{2}\Phi_{-}+
\left( g_{11}|\Phi _{-}|^{2}+g_{12}|\Phi _{+}|^{2}\right) \Phi_{-}%
+V(\mathbf{r})\,\Phi _{-}, \label{1} \\
i\partial _{t}{\Phi _{+}} &=&-\frac{1}{2}\nabla^{2}\Phi_{+}+
\left( g_{21}|\Phi _{-}|^{2}+g_{22}|\Phi_{+}|^{2}\right)\Phi_{+}%
+V(\mathbf{r})\,\Phi _{+},  \label{2}
\end{eqnarray}
\label{start_manakov}
\end{subequations}
where $\nabla^{2}$ stands for the standard Laplace operator in the
respective dimension of the problem, the interaction coefficients 
are $g_{jk}>0$ ($\forall j,k=1,2$), with $g_{21}\equiv g_{12}$, 
and the external potential $V(\mathbf{r})$ assumes the standard harmonic
form of $V(\mathbf{r})=\frac{1}{2}\Omega^{2}|\mathbf{r}|^{2}$, with 
$|\mathbf{r}|^{2}=x^2+y^2+z^2$ and normalized trap strength $\Omega$
(note that in 1D, $\nabla^{2}=\partial_x^2$, $V=\frac{1}{2}\Omega^{2}x^{2}$, and so on). 
The fields (representing the macroscopic wavefunctions in BECs~\cite{siambook})  
$\Phi_{\pm }=\Phi_{\pm }(\mathbf{r},t)$ in Eqs.~(\ref{1})-(\ref{2}) 
are assumed to carry the dark (with \textquotedblleft $-$" subscript)
and bright (with \textquotedblleft $+$" subscript) soliton components, 
respectively.

The starting point for our discussion below is the construction of stationary
solutions. Such stationary solutions to Eqs.~(\ref{1}) and (\ref{2}) with
chemical potentials $\mu_{\pm}$ are found by employing the well-known
ansatz,
\begin{equation}
\Phi_{\pm }(\mathbf{r},t)=\phi_{\pm }(\mathbf{r})\exp (-i\mu _{\pm }t),
\label{stat}
\end{equation}
where $\phi_{\pm}(\mathbf{r})$ stand for the steady states of the corresponding
solitary waveforms. Then, Eqs.~(\ref{1})-(\ref{2}) reduce to the coupled system of
stationary equations
\begin{subequations}
\begin{eqnarray}
\mu _{-}\phi _{-} &=&-\frac{1}{2}\nabla^{2}\phi_{-}+%
\left(g_{11}|\phi _{-}|^{2}+g_{12}|\phi _{+}|^{2}\right) \phi_{-}+V(\mathbf{r})\,\phi _{-},
\label{-} \\
\mu _{+}\phi _{+} &=&-\frac{1}{2}\nabla^{2}\phi_{+}+%
\left(g_{12}|\phi _{-}|^{2}+g_{22}|\phi _{+}|^{2}\right) \phi_{+}+V(\mathbf{r})\,\phi _{+}.
\label{+}
\end{eqnarray}
\label{stat_nlse}
\end{subequations}

A key point in our analysis is that, as is well known (see, e.g., 
Ref.~\cite{parkshin}), the Manakov model [cf. Eqs.~(\ref{start_manakov})
in 1D with $g_{ij}=1$ and without an external potential] is invariant 
under the action of the $SU(2)$ Lie-group. In fact, this result does not 
depend on the dimensionality of the system or the presence
of an external potential (with the constraint that it should be
the same for the two components), as long as $g_{ij}=1$. 
Indeed, let us first recall that a general matrix element of $SU(2)$ has
the form
\[ U=\left( \begin{array}{ccc}
\alpha & -\bar{\beta} \\
\beta & \bar{\alpha} \end{array} \right),\]
where bar denotes complex conjugate, and complex constants 
$\alpha$ and $\beta$ are such that
$|\alpha|^2+|\beta|^2=1$. Then, it can be shown that if the (pseudo-) spinor 
$(\Phi_{-},~\Phi_{+})^{T}$ is a solution of Eqs.~(\ref{start_manakov}), then,
\[ \left( \begin{array}{ccc}
\Phi_{-}'\\
\Phi_{+}' \end{array} \right)
\equiv U \left( \begin{array}{ccc}
\Phi_{-} \\
\Phi_{+} \end{array} \right)
= \left( \begin{array}{ccc}
\alpha \Phi_{-}-\bar{\beta} \Phi_{+}\\
\beta \Phi_{-}+\bar{\alpha} \Phi_{+} \end{array} \right),
\]
is also a solution of Eqs.~(\ref{start_manakov}). In our 
considerations for what follows, we will focus on the special
case of an $SO(2)$ rotation parametrized by an angle $\delta\in[0,2\pi)$ with a
$2\times2$ matrix representation
\begin{equation}
U\equiv R(\delta)=
\begin{pmatrix}
\cos{\delta} & -\sin{\delta} \\
\sin{\delta} & \cos{\delta}
\end{pmatrix}
,
\label{so2_group}
\end{equation}
corresponding to the choice of $\alpha=\cos{\delta}$ and $\beta=\sin{\delta}$.
Then, once stationary solutions in the form of Eq.~(\ref{stat}) are identified, 
the rotation operator $R(\delta)$ given by Eq.~(\ref{so2_group}) acts
on $\vect{\Phi}=(\Phi_{-},\Phi_{+})^T$ as follows:
\begin{equation}
\vect{\Phi} \rightarrow \vect{\Phi'} = R(\delta)\, \vect{\Phi} =
\begin{pmatrix}
\cos{\delta}\,\phi_{-}\exp(-i\mu_{-}t)-\sin{\delta}\,\phi_{+}\exp(-i\mu_{+}t)\\
\sin{\delta}\,\phi_{-}\exp(-i\mu_{-}t)+\cos{\delta}\,\phi_{+}\exp(-i\mu_{+}t)
\end{pmatrix}
.
\label{so2_rot}
\end{equation}
It is now straightforward to determine the densities of the rotated fields
$\Phi_{\pm}'$, which read:
\begin{subequations}
\begin{eqnarray}
n_{-}' &\equiv& |\Phi_{-}'|^2= |\phi_{-}|^2 \cos^2 \delta + |\phi_{+}|^2 \sin^2 \delta 
- \sin(2\delta) {\rm Re}\{\phi_{+}\bar{\phi}_{-}\exp[i\,\Delta\mu\,t]\}, \\
n_{+}' &\equiv& |\Phi_{-}'|^2= |\phi_{-}|^2 \sin^2 \delta + |\phi_{+}|^2 \cos^2 \delta 
+ \sin(2\delta) {\rm Re}\{\phi_{+}\bar{\phi}_{-}\exp[i\,\Delta\mu\,t)\},
\end{eqnarray}
\end{subequations}
where $\Delta \mu=\mu_{-}-\mu_{+}$. The above equations indicate that the
total density, 
\begin{equation}
n'=n_{-}'+n_{+}'= |\phi_{-}|^2 + |\phi_{+}|^2, 
\end{equation}
is time-independent (recall that $\phi_{\pm}$ depend only on $\mathbf{r}$), 
while the individual densities $n_{\pm}'$ of the rotated states are periodic
functions of time. In fact, the relevant angular frequency, which constitutes
the internal beating frequency of the rotated structures, is $\omega=\Delta \mu$, while 
the period of internal vibrations is given by:
\begin{equation}
T=\frac{2\pi}{\Delta \mu}.
\label{period}
\end{equation}

Our algorithm for the construction of the rotated (beating) 
dark-dark solitons and generalizations thereof in higher dimensions
is described as follows. At first, we identify
steady states $\phi_{\pm}$ to Eq.~(\ref{stat_nlse}) using a Newton-Raphson
method for a given set of chemical potentials $\mu_{\pm}$.
This state will be of the dark-bright variety in 1D, of the
vortex-bright in 2D, and of the vortex-line (VL)--bright (VLB)
and vortex-ring (VR)--bright (VRB) in 3D.
In the majority of the cases studied below, we consider two
cases as far as the interaction coefficients are concerned (unless
otherwise noted): 
\begin{itemize}
\item[(i)] $g_{ij}=1$, i.e., equal interaction coefficients, and 
\item[(ii)] unequal ones with $g_{11}=1.03$, $g_{12}=1$, and $g_{22}=0.97$.
\end{itemize}
We have used these values as ``typical'' ones appearing in the context 
of $^{87}$Rb BECs \cite{originalhall}, although the precise value of 
the coefficients is still under active investigation; see, e.g., 
the discussion of Ref.~\cite{refhall} and
references therein.

Subsequently, the steady states obtained numerically are transformed by
utilizing the orthogonal transformation given by Eq.~(\ref{so2_rot}), 
where we only consider the cases with $\delta=\pi/4$ and $\delta=\pi/8$.
Then, having the rotated waveforms at hand, we supply them (at $t=0$) as
initial conditions, and advance Eq.~(\ref{start_manakov}) forward in time
using a standard fourth-order Runge-Kutta method (RK4) and its parallel
version (using OpenMP) with fixed time-step. We refer the interested reader
to Refs.~\cite{dbs_coupled_Manakov,vbs_coupled,dsss_ww} for a detailed description
on the numerical methods employed in this present work. In our numerical
computations presented below, we consider values of the trap strength
$\Omega$ of $0.1, 0.2$ and $1$ in the 1D, 2D and 3D cases, respectively.
In our one- and two-dimensional settings, we also explore the scenarios
in the absence of a trap (i.e., for $\Omega=0$).

We should also notice that in the Manakov
case where the transformation is exact, 
the stability of the rotated states is inherited
from their stationary counterparts and, consequently, all the 
dynamical solutions considered are stable. On the other hand,
for the case with $g_{ij} \neq 1$, the situation may be  more subtle
as will be explained in more detail through our numerical results below.

\section{Numerical Results}

\subsection{Dark-dark solitons in 1D}

We start by considering, at first, the 1D case. 
It is relevant to mention that while corresponding analysis
has been presented earlier, e.g., in Refs.~\cite{pe4,pe5}
(see also~\cite{parkshin}), we provide the relevant case
examples in order to set the stage for our higher-dimensional
generalizations.

\begin{figure}[t]
\begin{center}
\vspace{-0.1cm}
\mbox{\hspace{-0.1cm}
\subfigure[][]{\hspace{-0.3cm}
\includegraphics[height=.16\textheight, angle =0]{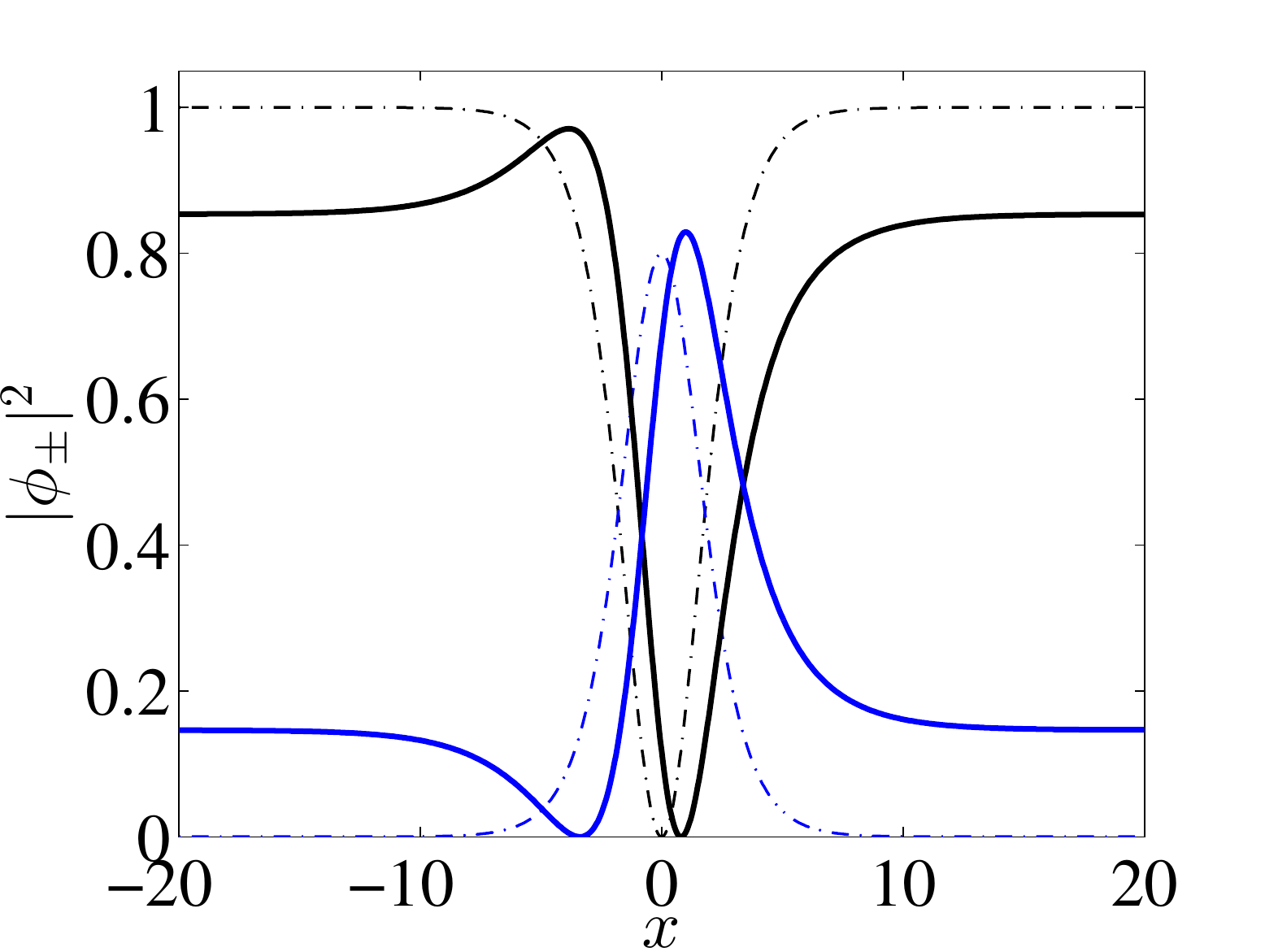}
\label{fig1a}
}
\subfigure[][]{\hspace{-0.3cm}
\includegraphics[height=.16\textheight, angle =0]{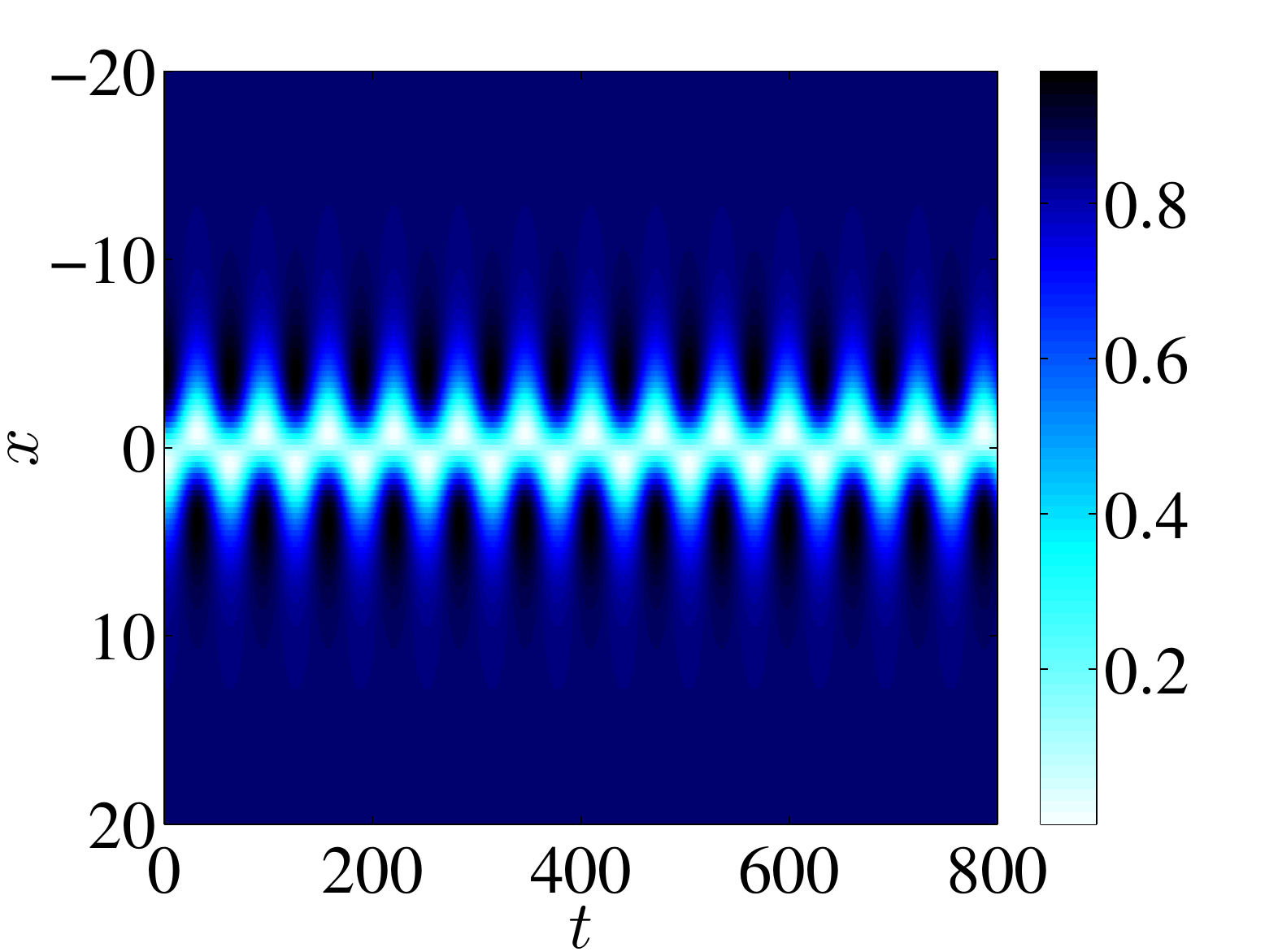}
\label{fig1b}
}
}
\mbox{\hspace{-0.1cm}
\subfigure[][]{\hspace{-0.3cm}
\includegraphics[height=.16\textheight, angle =0]{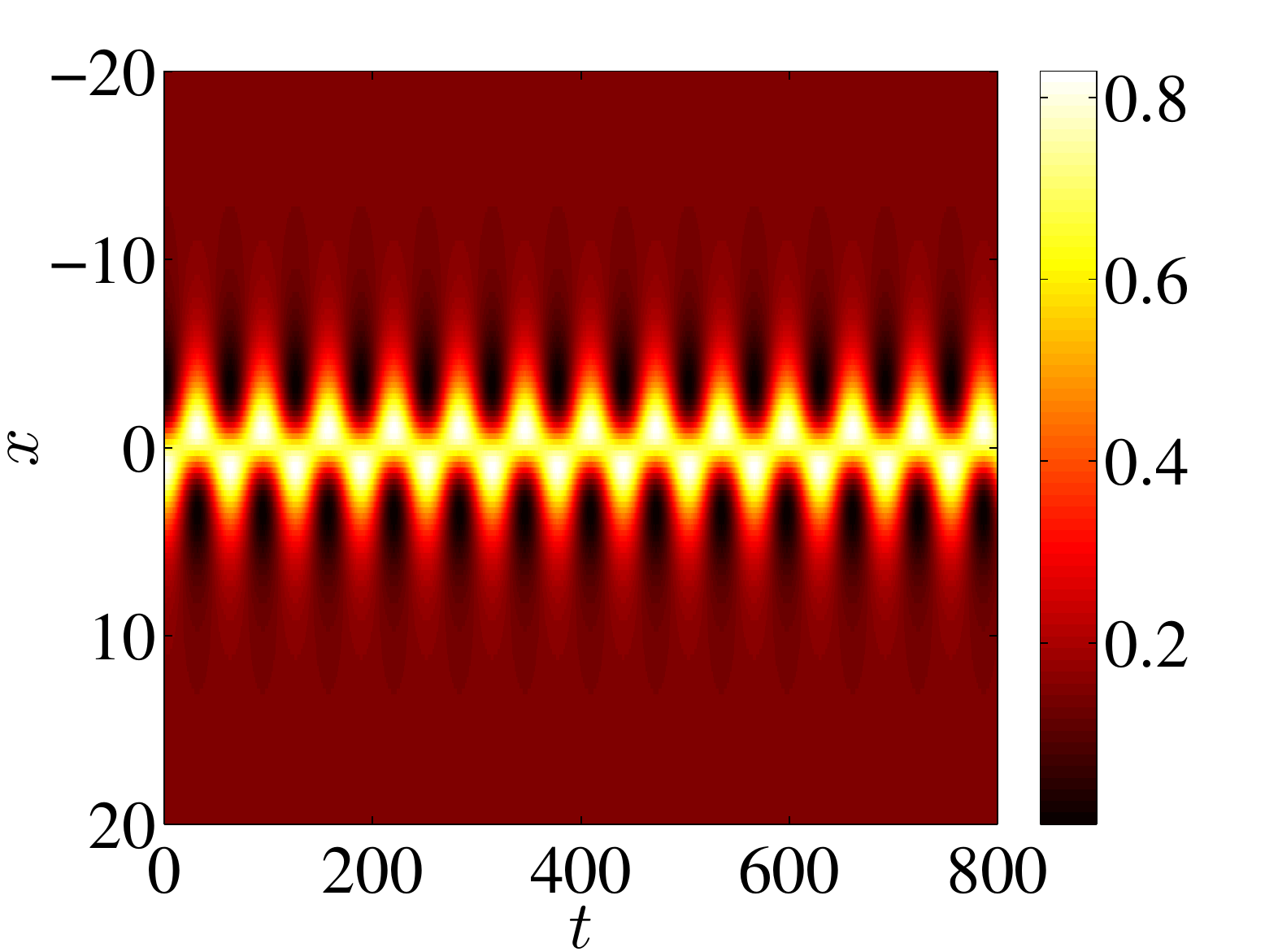}
\label{fig1c}
}
}
\mbox{\hspace{-0.1cm}
\subfigure[][]{\hspace{-0.3cm}
\includegraphics[height=.16\textheight, angle =0]{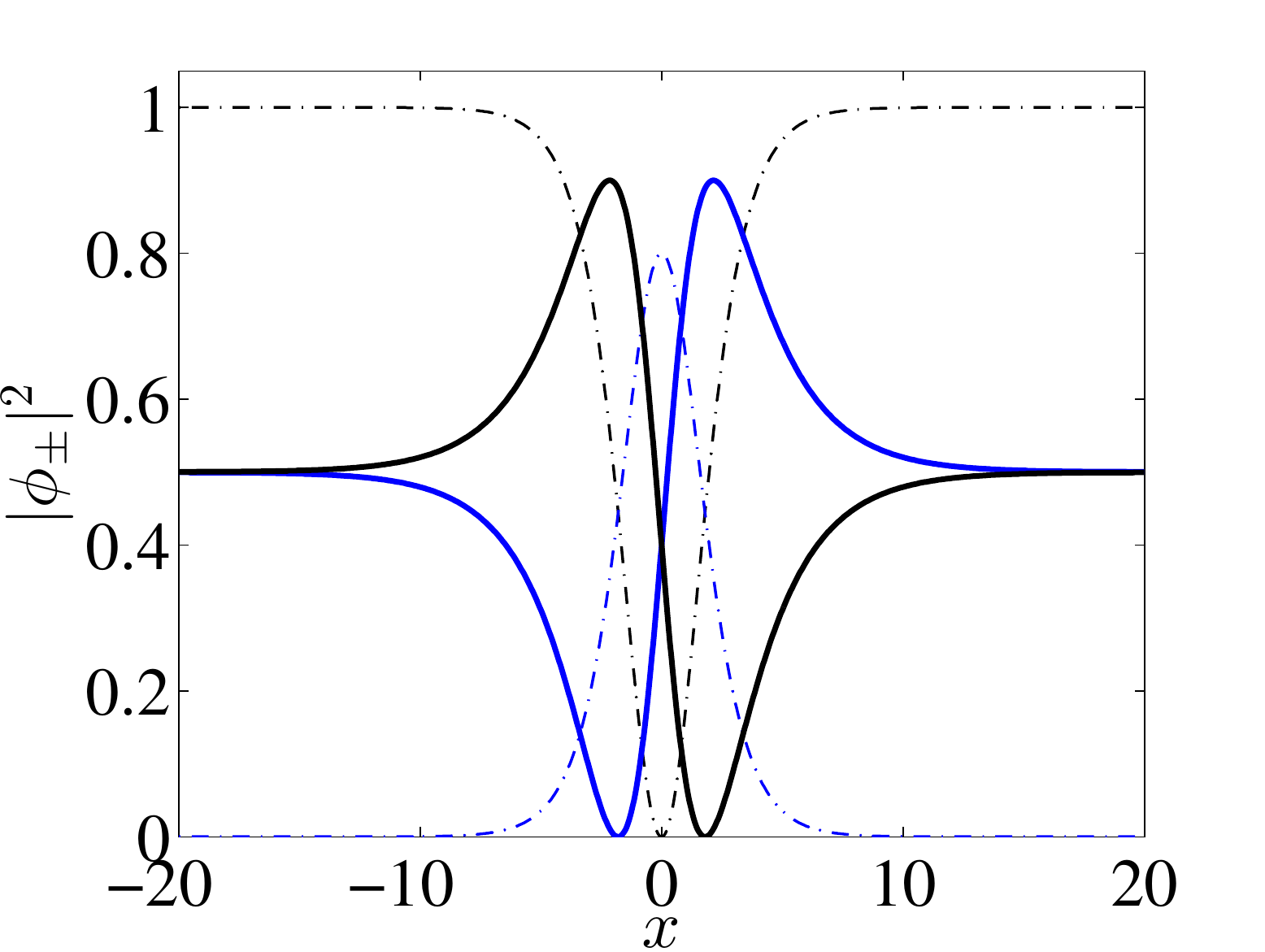}
\label{fig1d}
}
\subfigure[][]{\hspace{-0.3cm}
\includegraphics[height=.16\textheight, angle =0]{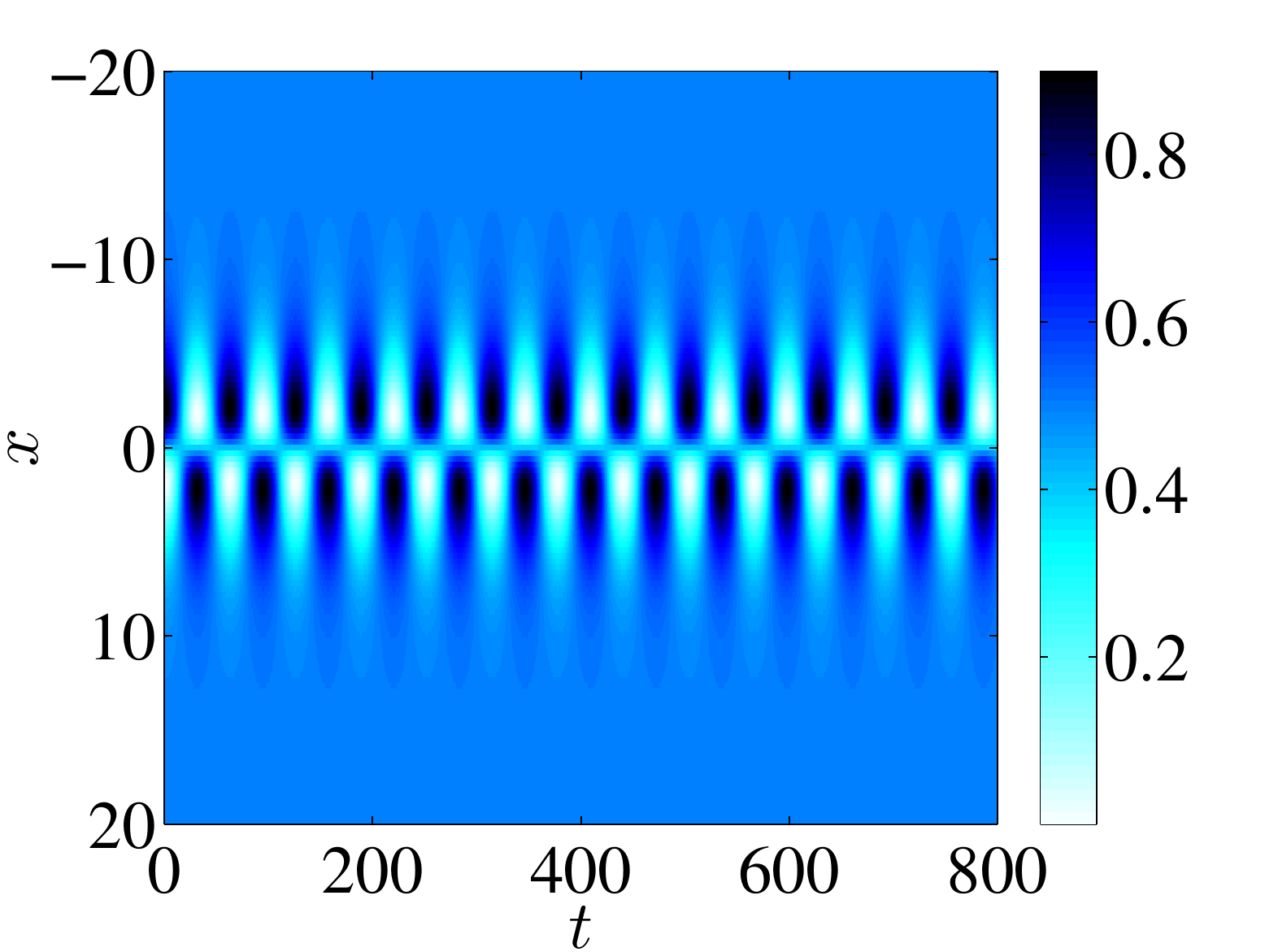}
\label{fig1e}
}
}
\mbox{\hspace{-0.1cm}
\subfigure[][]{\hspace{-0.3cm}
\includegraphics[height=.16\textheight, angle =0]{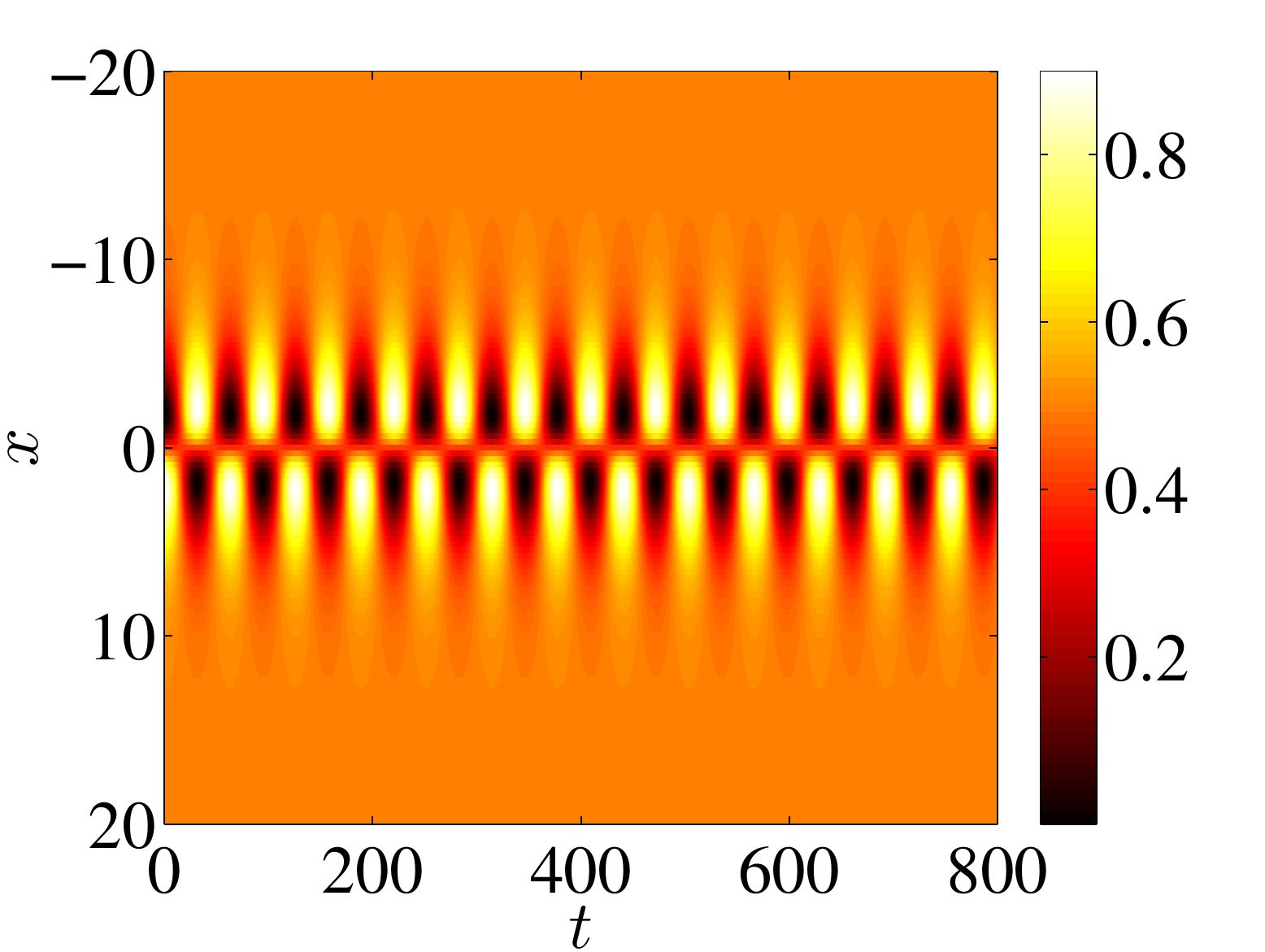}
\label{fig1f}
}
}
\mbox{\hspace{-0.1cm}
\subfigure[][]{\hspace{-0.3cm}
\includegraphics[height=.16\textheight, angle =0]{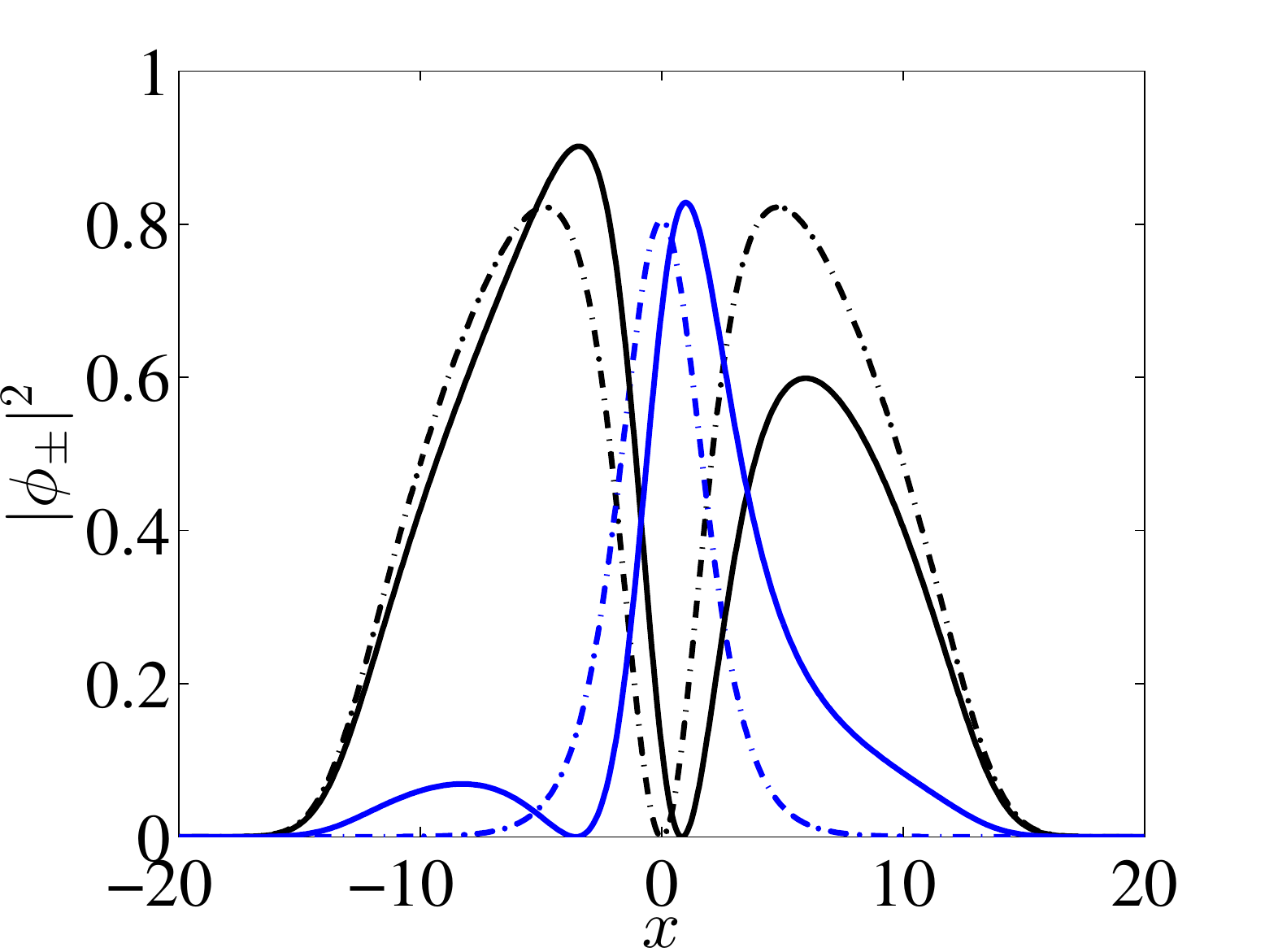}
\label{fig1g}
}
\subfigure[][]{\hspace{-0.3cm}
\includegraphics[height=.16\textheight, angle =0]{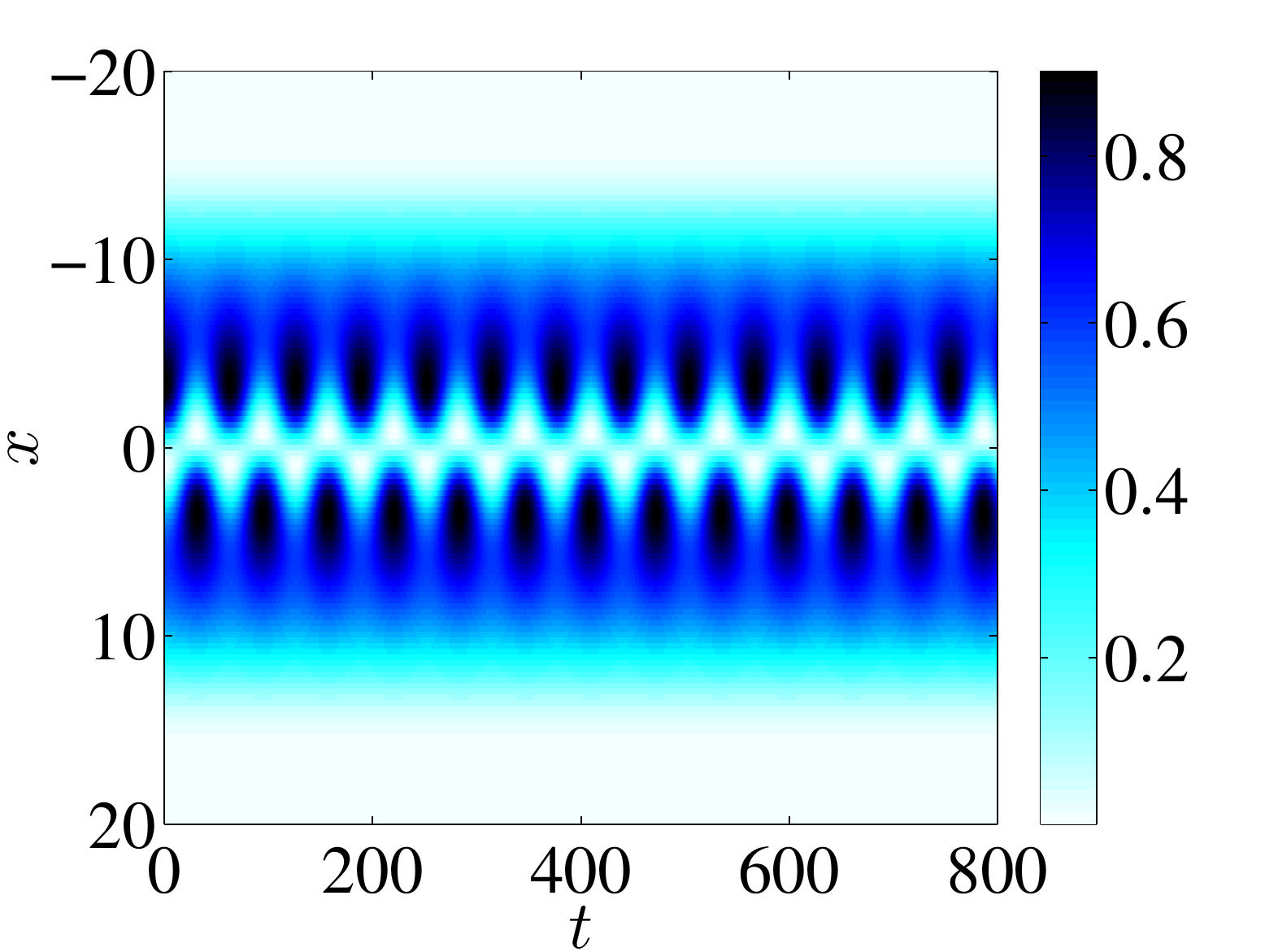}
\label{fig1h}
}
}
\mbox{\hspace{-0.1cm}
\subfigure[][]{\hspace{-0.3cm}
\includegraphics[height=.16\textheight, angle =0]{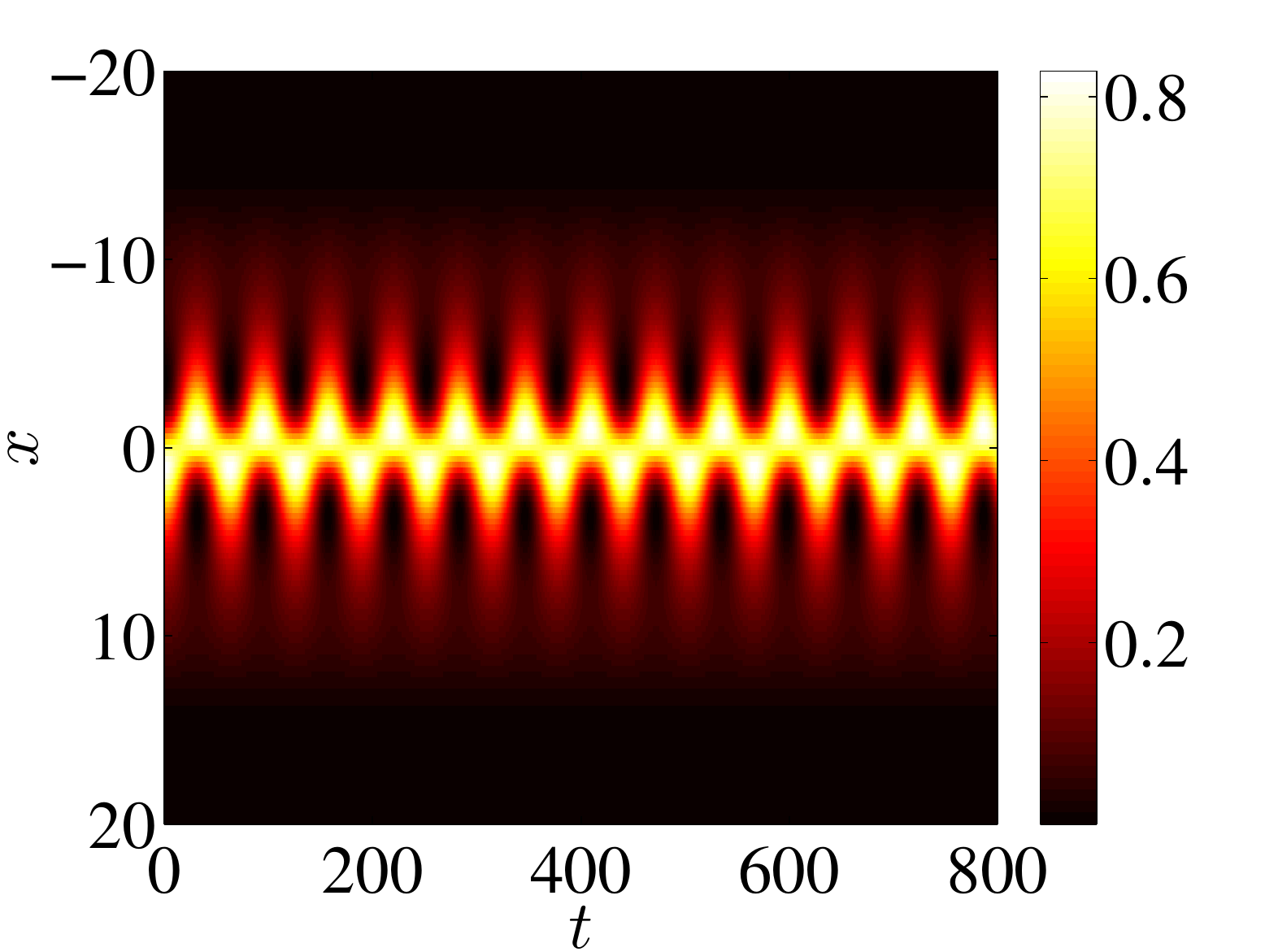}
\label{fig1i}
}
}
\mbox{\hspace{-0.1cm}
\subfigure[][]{\hspace{-0.3cm}
\includegraphics[height=.16\textheight, angle =0]{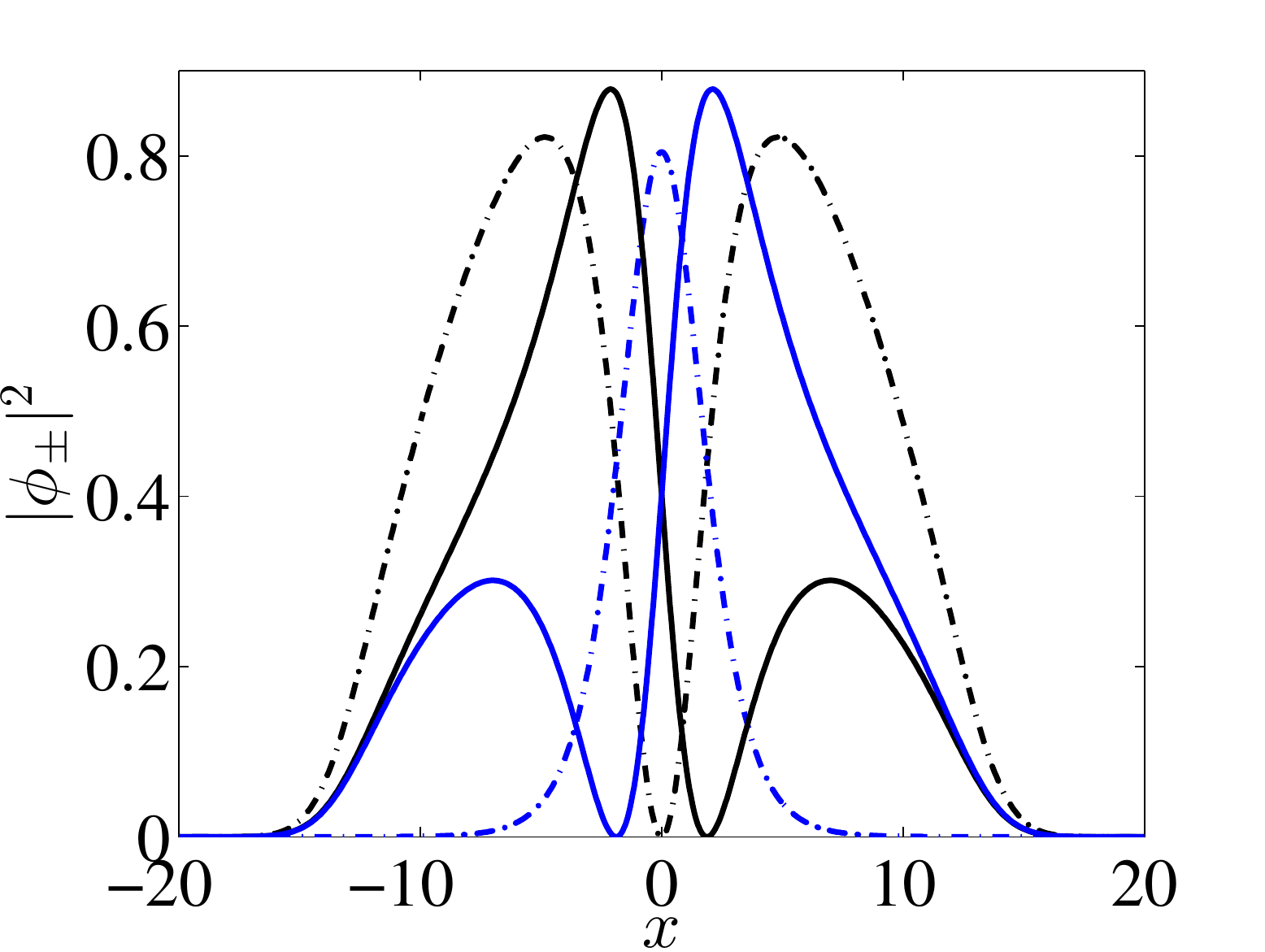}
\label{fig1j}
}
\subfigure[][]{\hspace{-0.3cm}
\includegraphics[height=.16\textheight, angle =0]{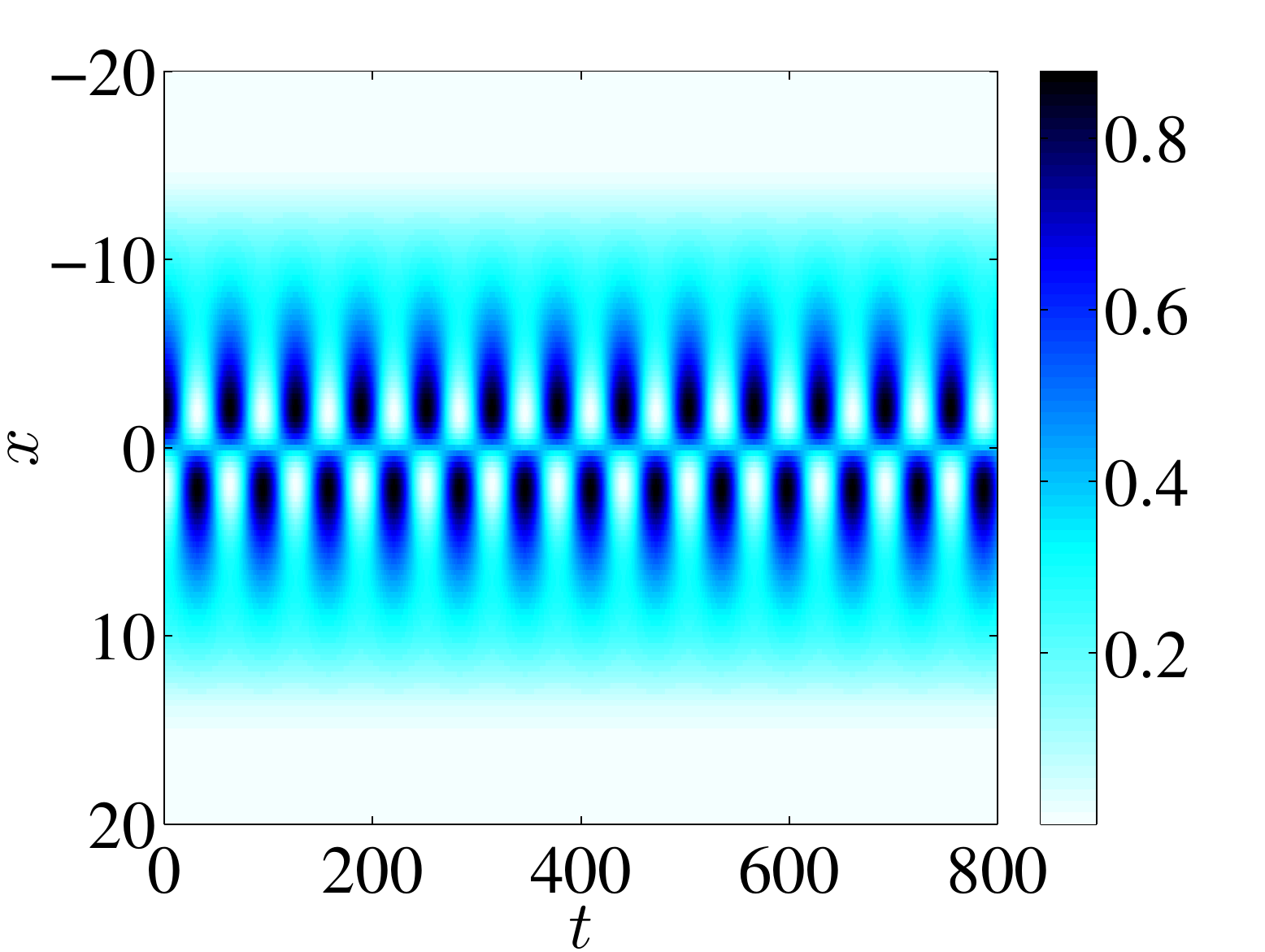}
\label{fig1k}
}
}
\mbox{\hspace{-0.1cm}
\subfigure[][]{\hspace{-0.3cm}
\includegraphics[height=.16\textheight, angle =0]{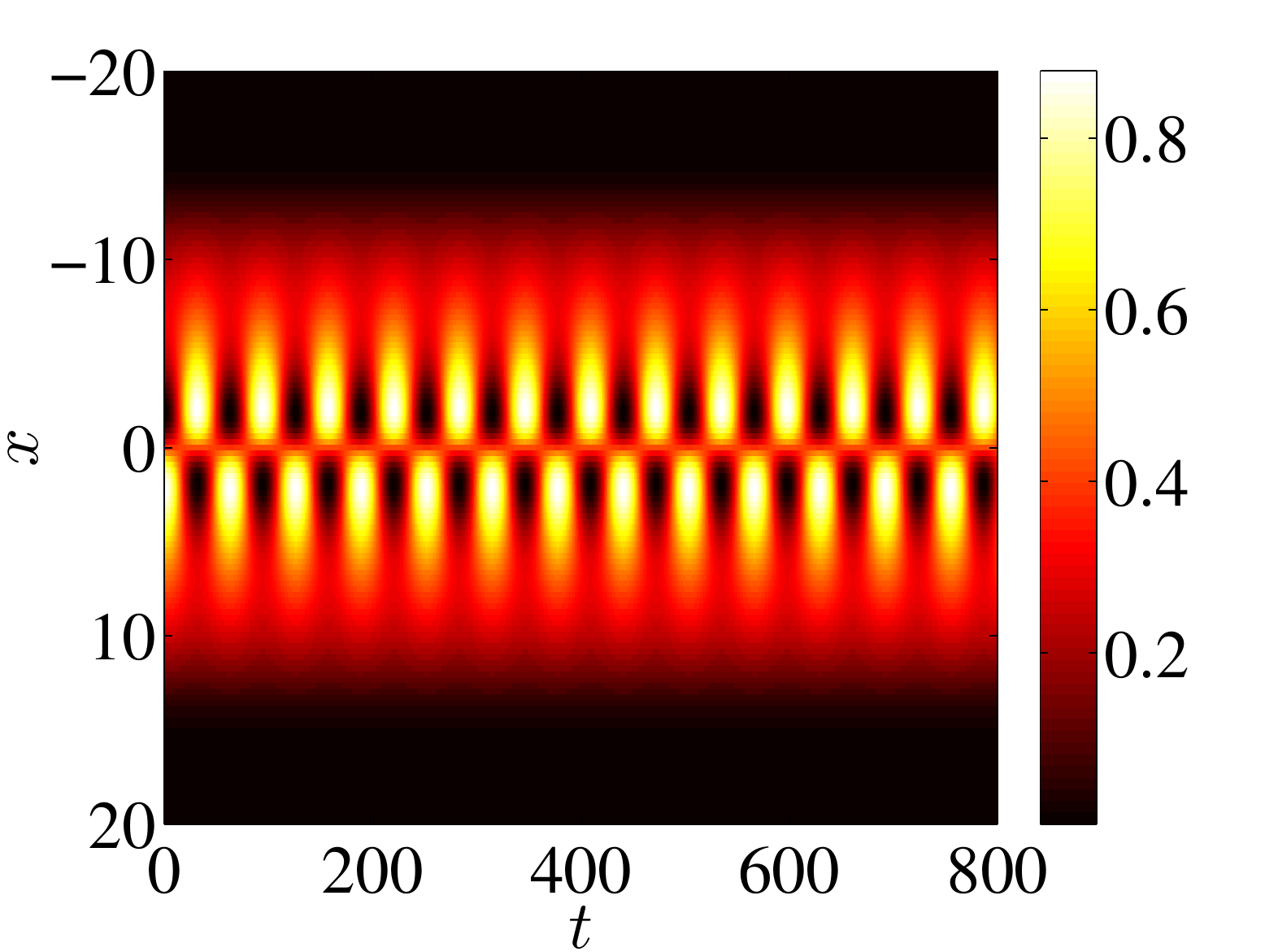}
\label{fig1l}
}
}
\end{center}
\par
\vspace{-0.7cm}
\caption{
(Color online) Summary of results corresponding to the case
with equal interaction coefficients and values of the chemical
potentials of $\mu_{-}=1$ and $\mu_{+}=0.9$. Panels (a)-(c), (g)-(i),
and (d)-(f), (j)-(l) correspond to a rotation of the original
steady state by $\delta=\pi/8$ and $\delta=\pi/4$, respectively.
In addition, the first two rows correspond to the homogeneous case
whereas the last two are shown in the presence of an external potential with
trap strength of $\Omega=0.1$. The left column presents the corresponding
$SO(2)$-rotated waveforms at $t=0$ for each case depicted by solid
blue (for the bright component) and black (for the dark one) lines.
Also, the original unrotated dark (dash-dotted black line) and bright
(dash-dotted blue line) solitary waveforms are depicted as well for
comparison. The spatio-temporal evolution of the densities $|\Phi_{-}(x,t)|^{2}$
and $|\Phi_{+}(x,t)|^{2}$ is presented in the middle and right
columns, respectively, with different colormaps in order to differentiate
between the two.
}
\label{fig1}
\end{figure}

\begin{figure}[t]
\begin{center}
\vspace{-0.1cm}
\mbox{\hspace{-0.1cm}
\subfigure[][]{\hspace{-0.3cm}
\includegraphics[height=.15\textheight, angle =0]{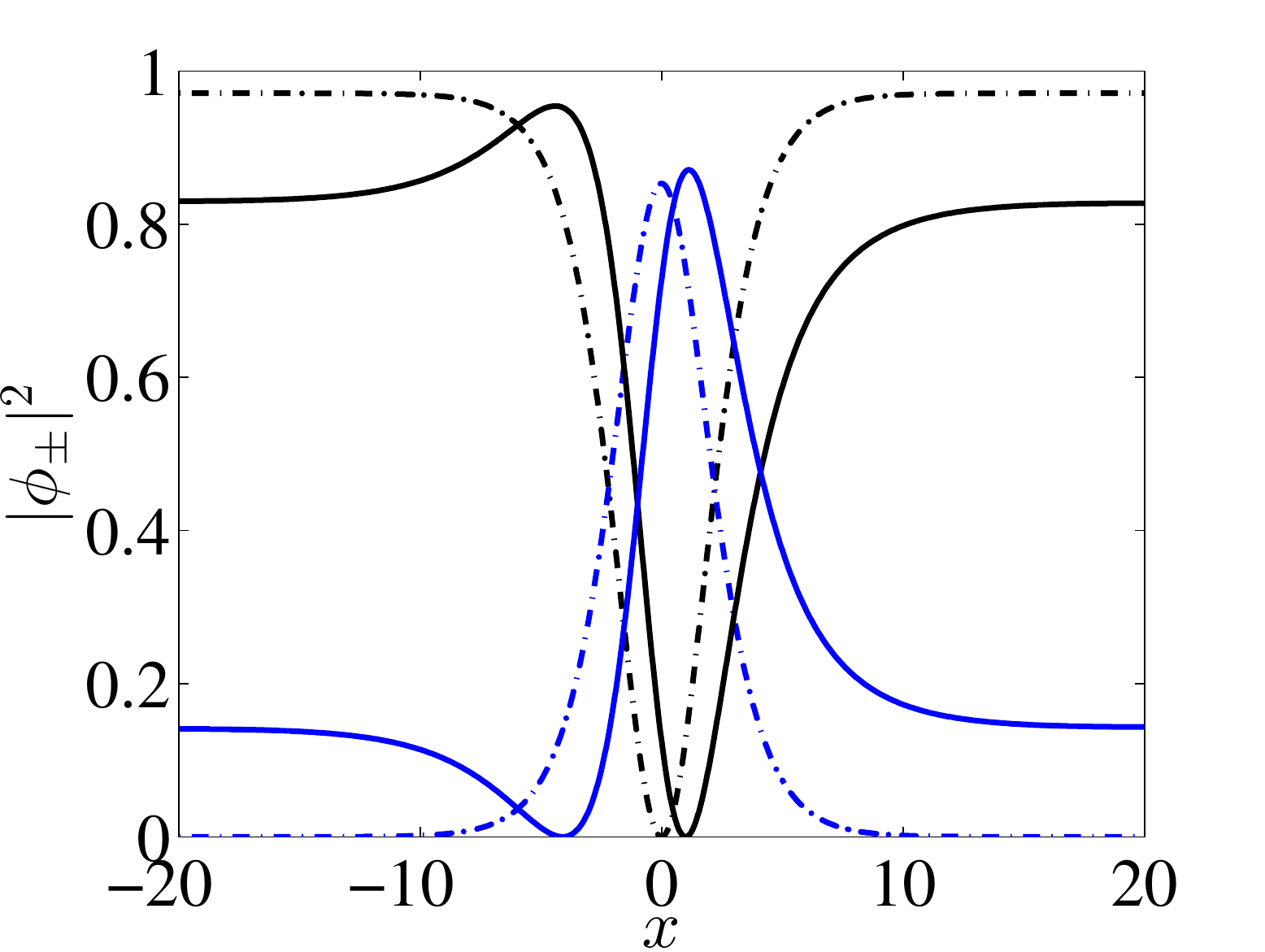}
\label{fig2a}
}
\subfigure[][]{\hspace{-0.3cm}
\includegraphics[height=.15\textheight, angle =0]{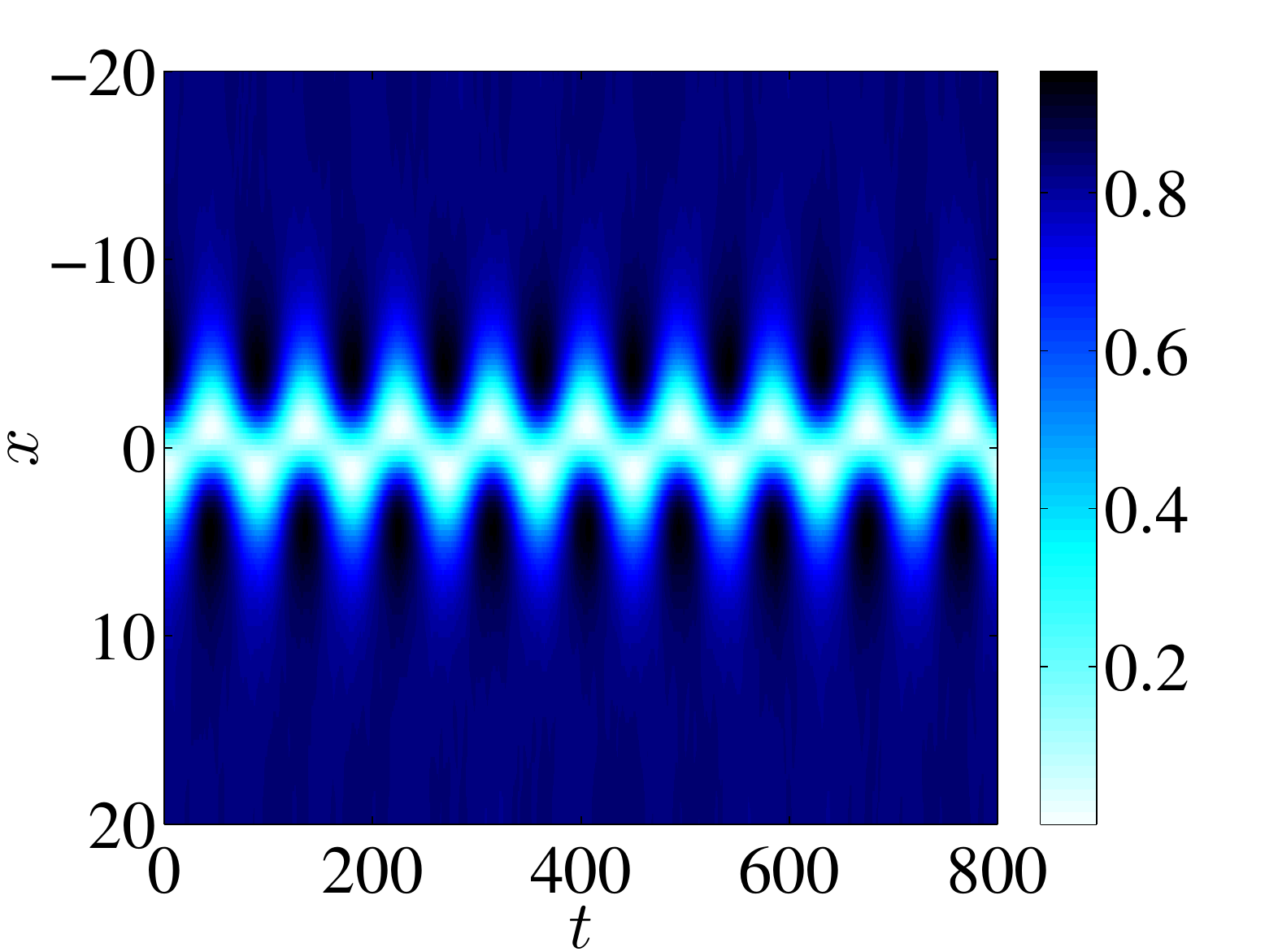}
\label{fig2b}
}
}
\mbox{\hspace{-0.1cm}
\subfigure[][]{\hspace{-0.3cm}
\includegraphics[height=.15\textheight, angle =0]{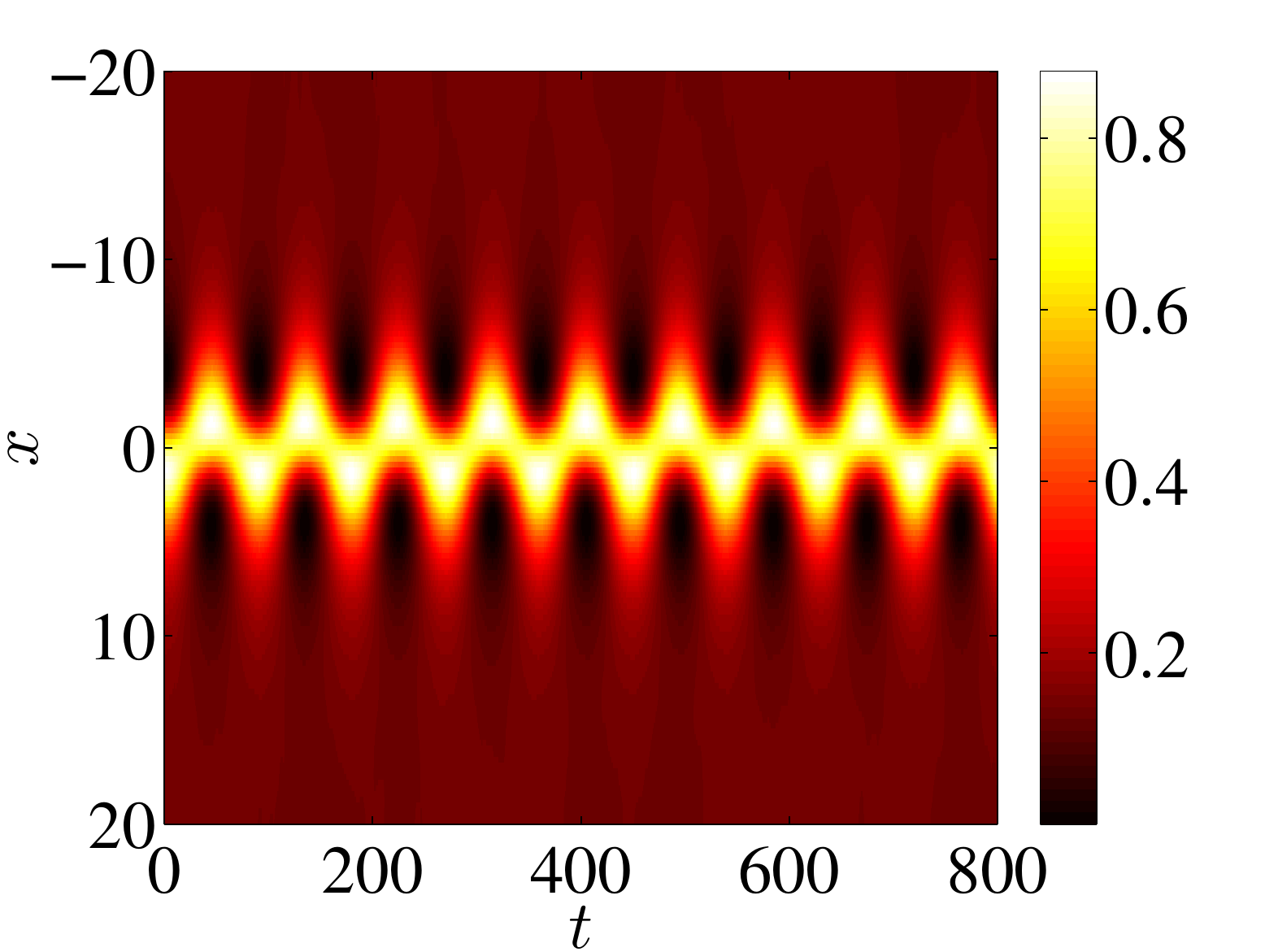}
\label{fig2c}
}
}
\mbox{\hspace{-0.1cm}
\subfigure[][]{\hspace{-0.3cm}
\includegraphics[height=.15\textheight, angle =0]{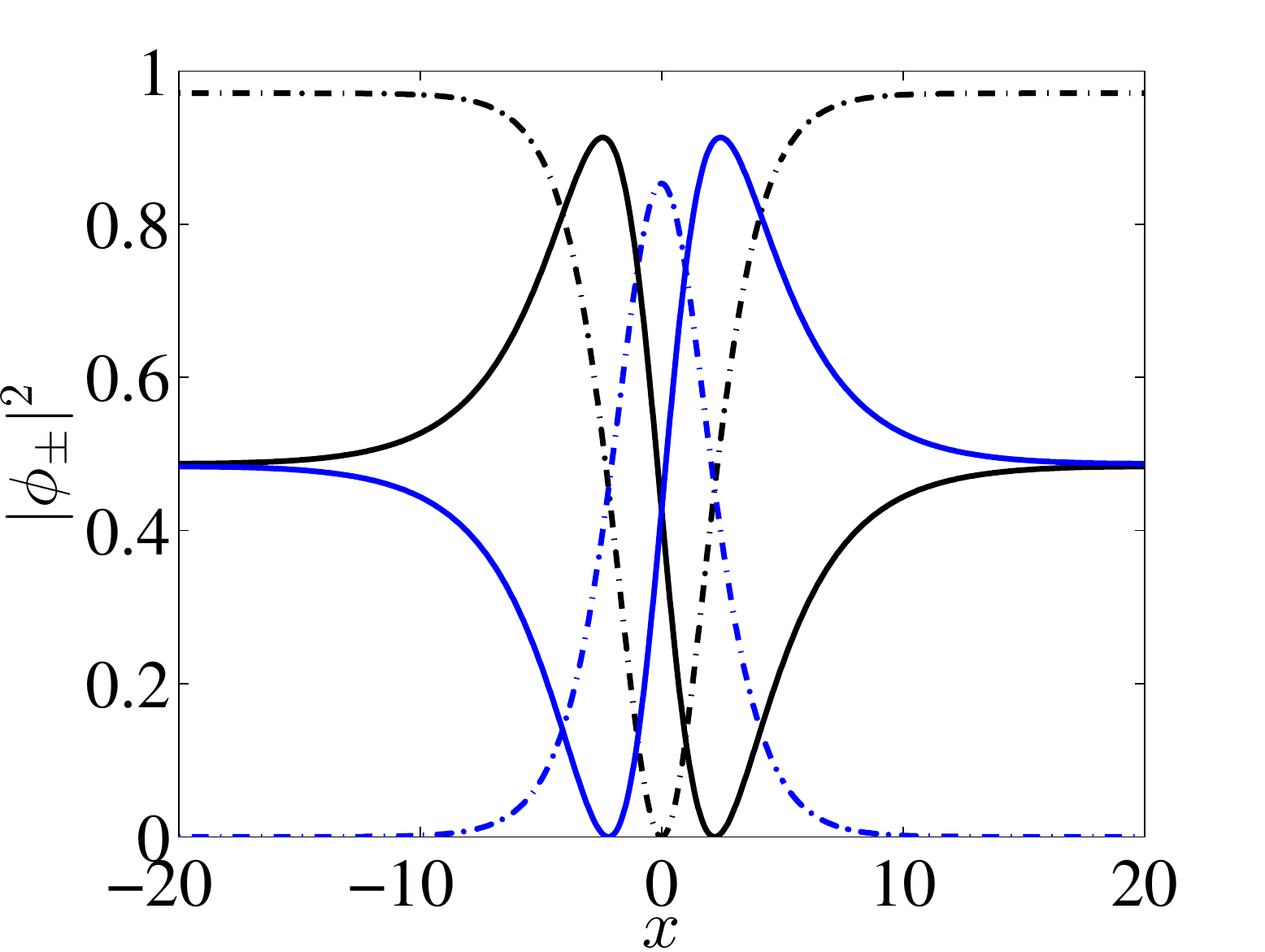}
\label{fig2d}
}
\subfigure[][]{\hspace{-0.3cm}
\includegraphics[height=.15\textheight, angle =0]{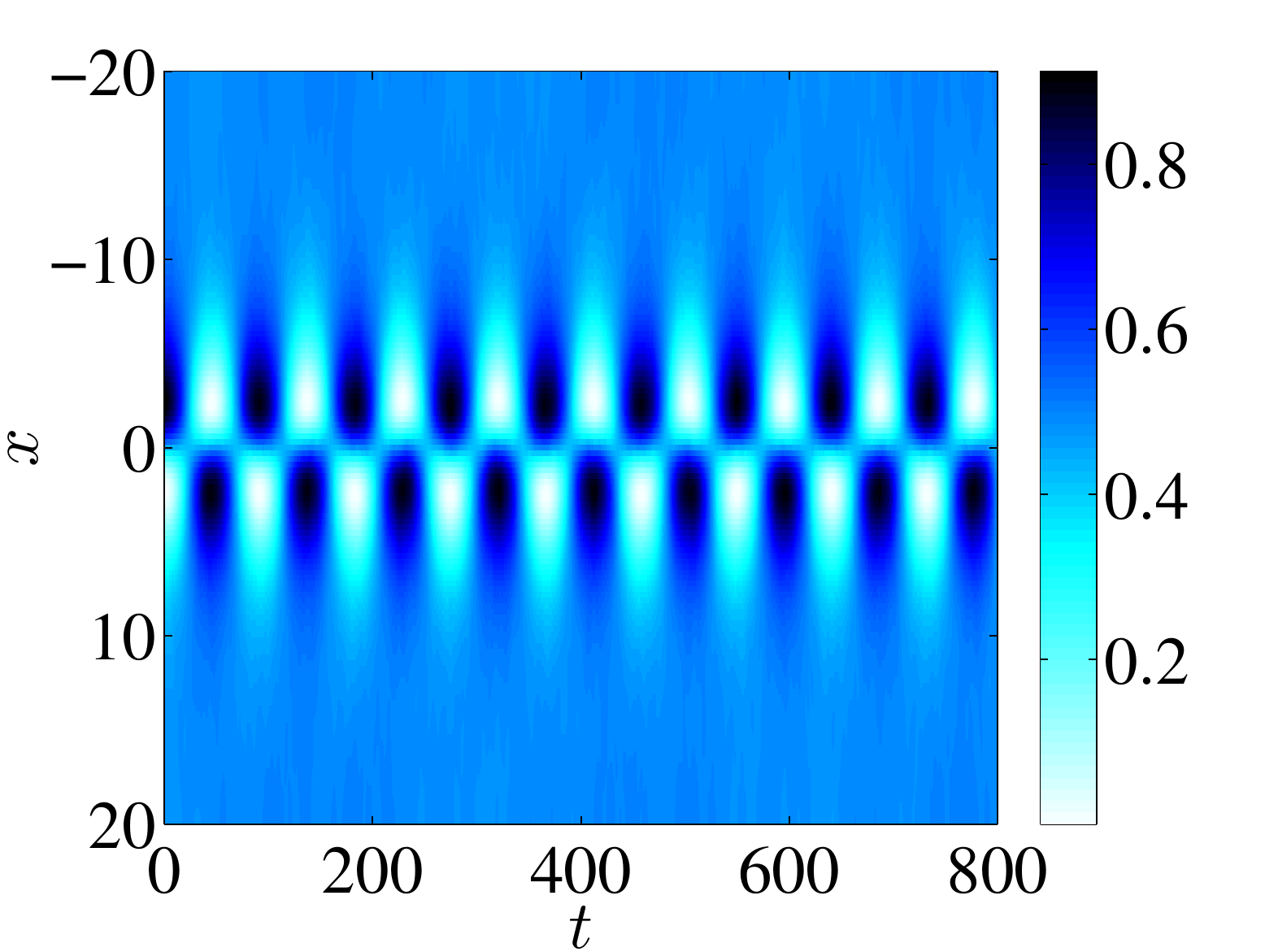}
\label{fig2e}
}
}
\mbox{\hspace{-0.1cm}
\subfigure[][]{\hspace{-0.3cm}
\includegraphics[height=.15\textheight, angle =0]{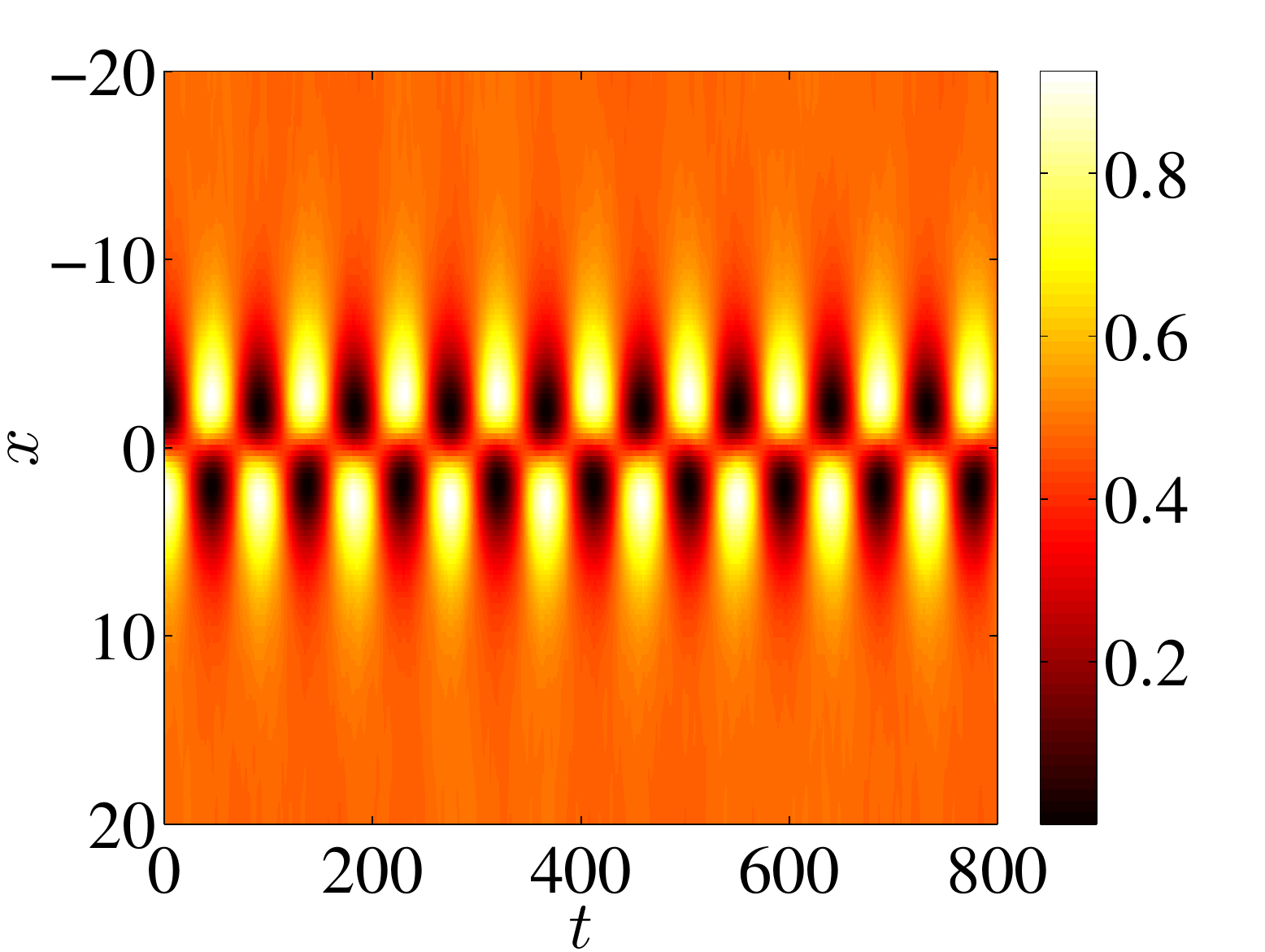}
\label{fig2f}
}
}
\mbox{\hspace{-0.1cm}
\subfigure[][]{\hspace{-0.3cm}
\includegraphics[height=.15\textheight, angle =0]{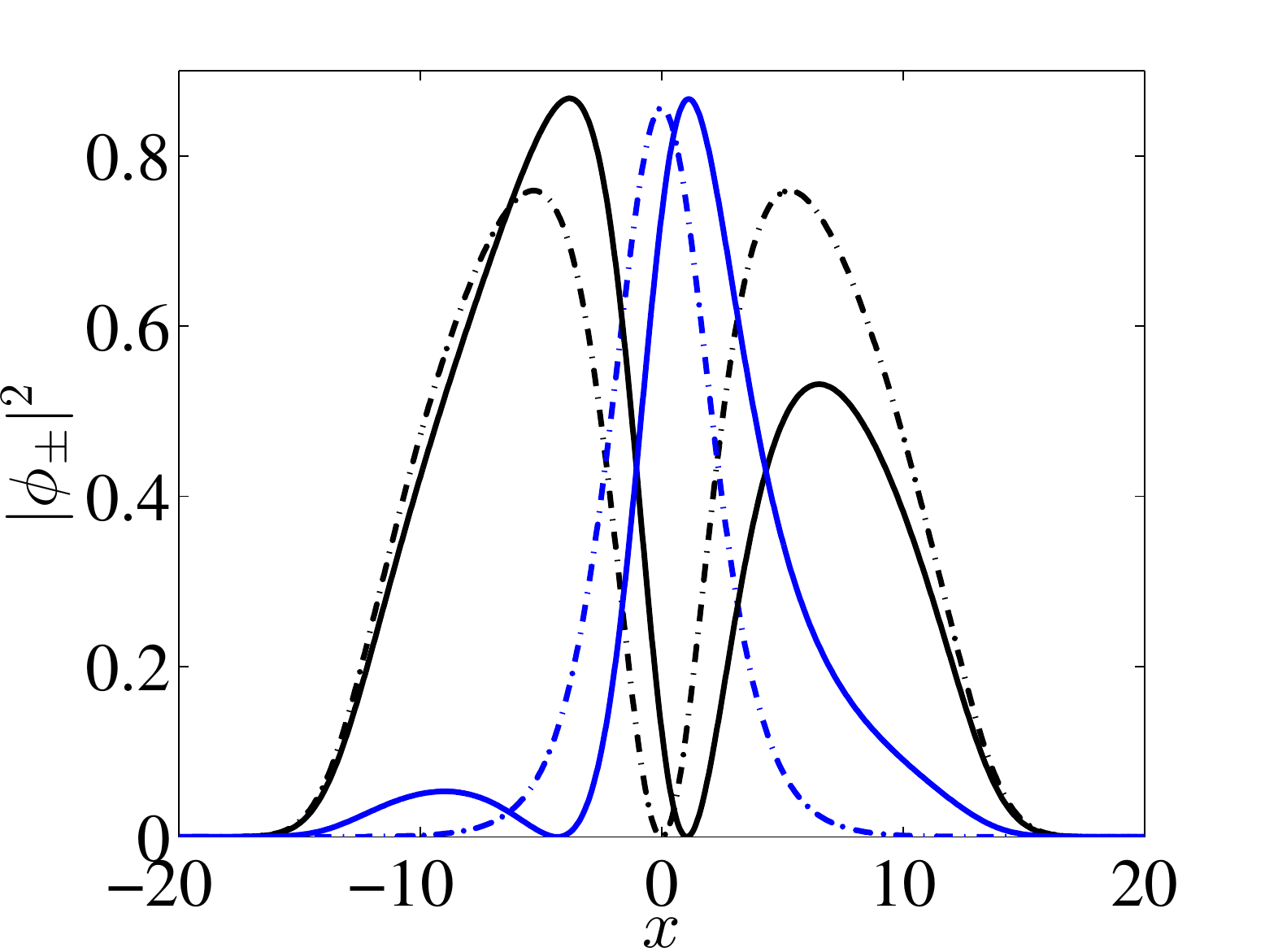}
\label{fig2g}
}
\subfigure[][]{\hspace{-0.3cm}
\includegraphics[height=.15\textheight, angle =0]{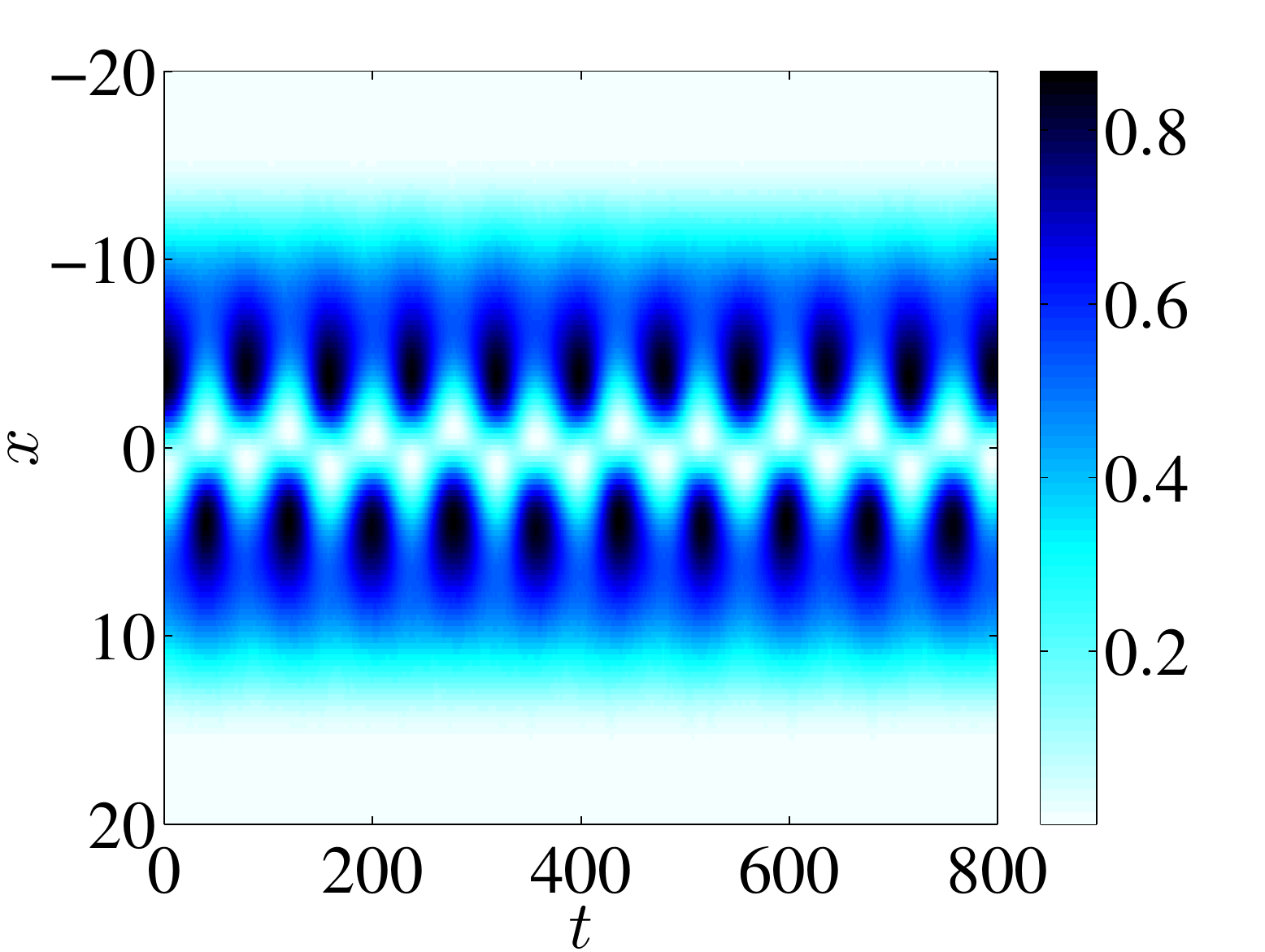}
\label{fig2h}
}
}
\mbox{\hspace{-0.1cm}
\subfigure[][]{\hspace{-0.3cm}
\includegraphics[height=.15\textheight, angle =0]{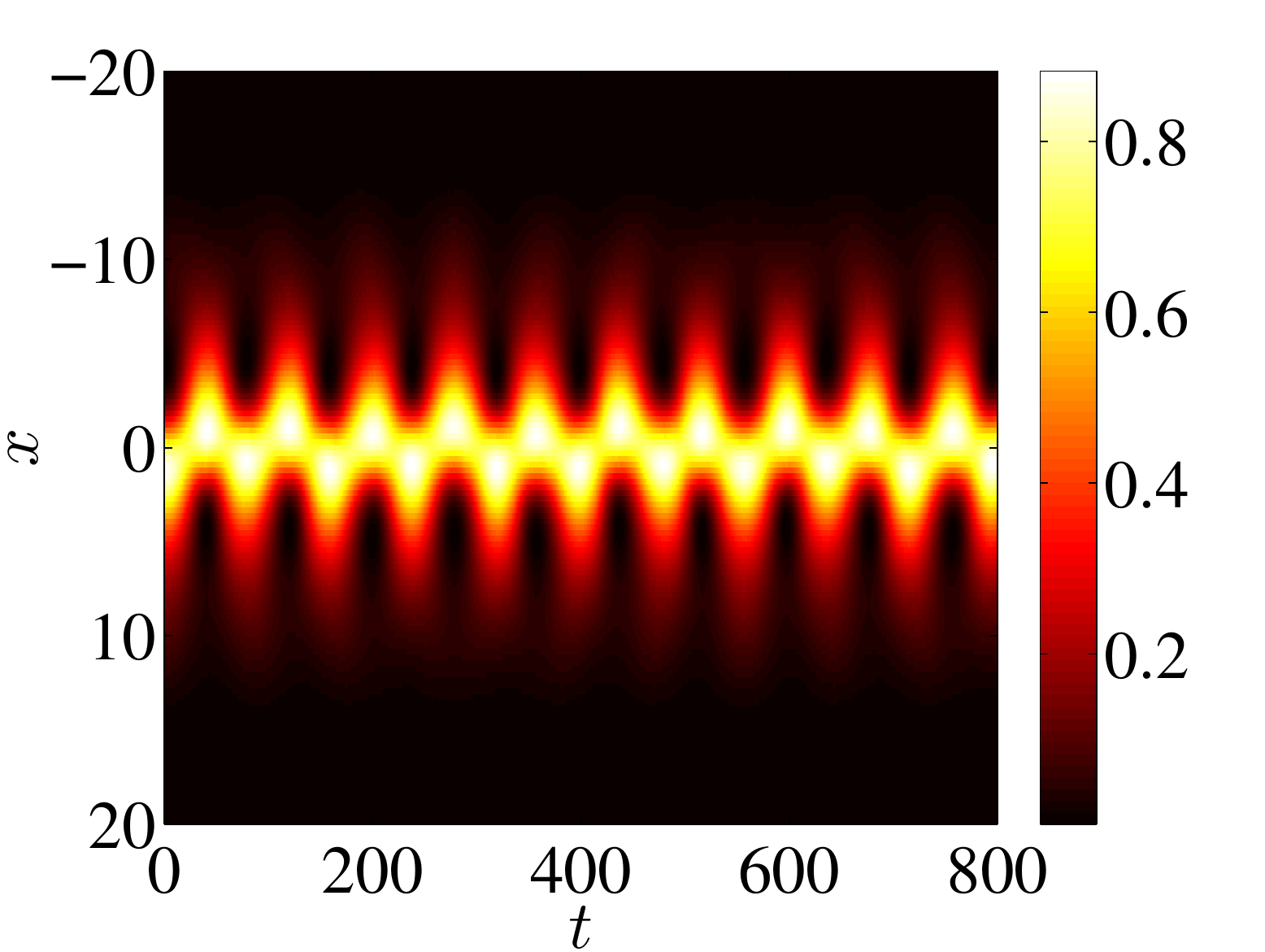}
\label{fig2i}
}
}
\mbox{\hspace{-0.1cm}
\subfigure[][]{\hspace{-0.3cm}
\includegraphics[height=.15\textheight, angle =0]{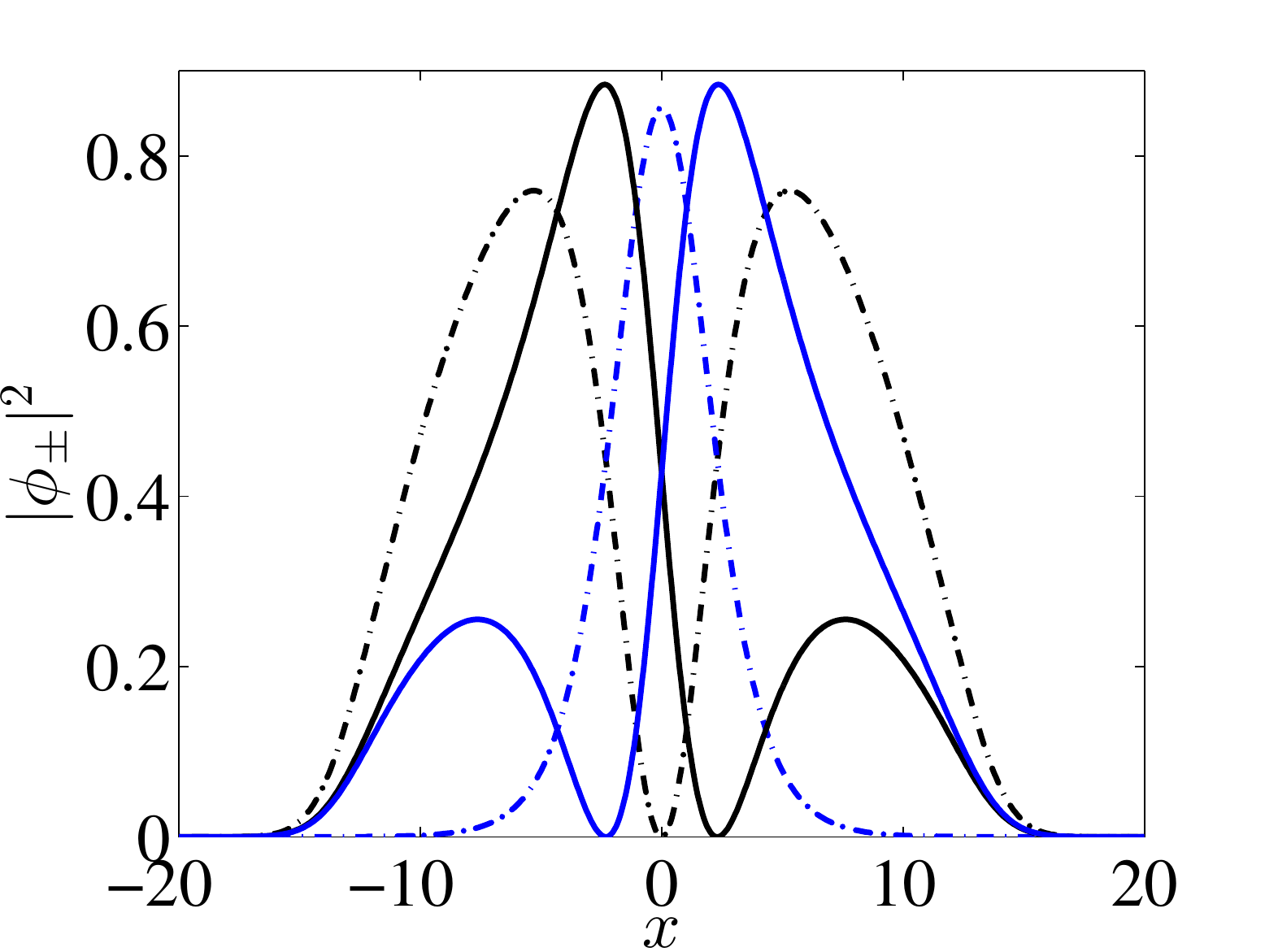}
\label{fig2j}
}
\subfigure[][]{\hspace{-0.3cm}
\includegraphics[height=.15\textheight, angle =0]{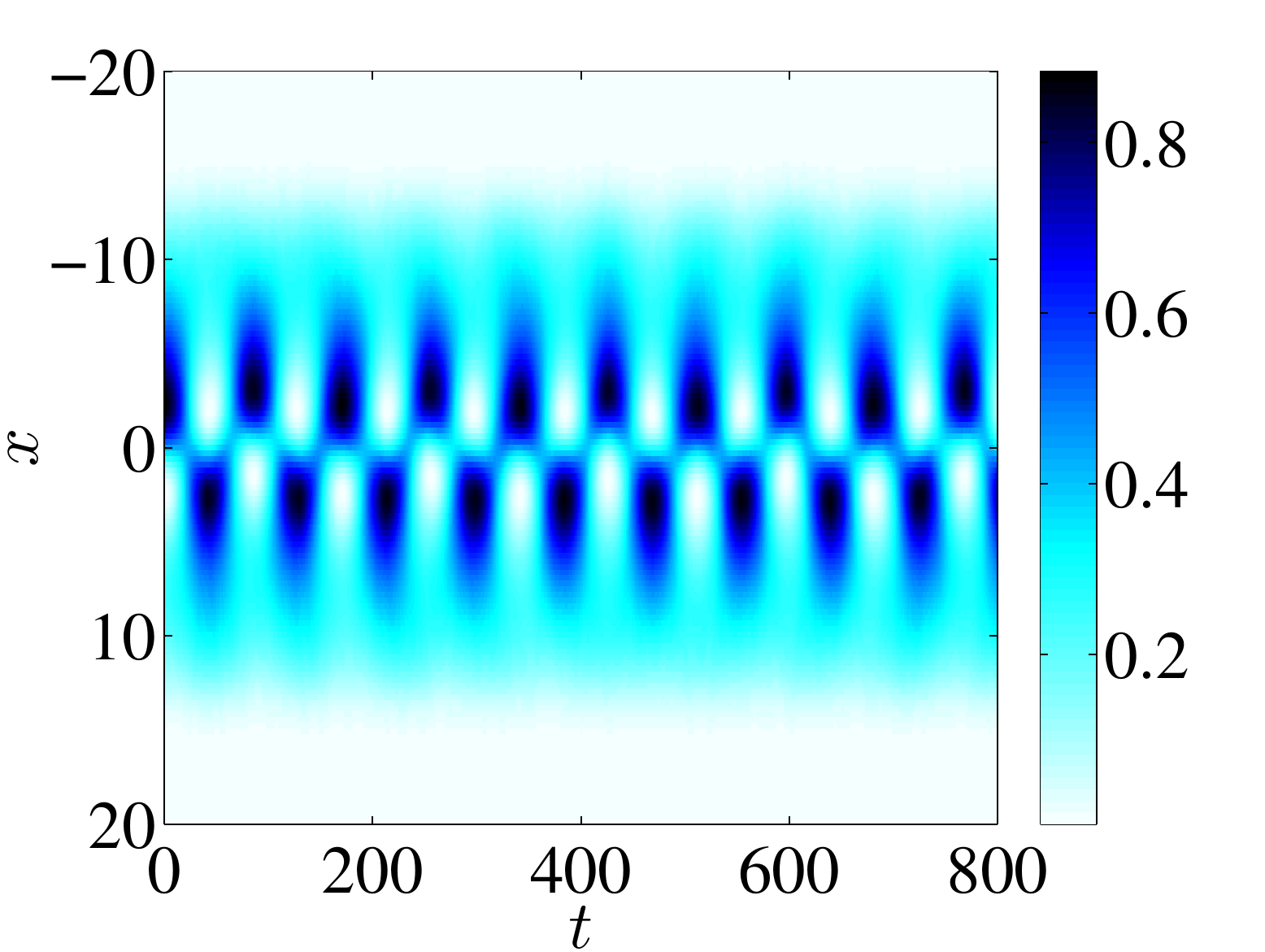}
\label{fig2k}
}
}
\mbox{\hspace{-0.1cm}
\subfigure[][]{\hspace{-0.3cm}
\includegraphics[height=.15\textheight, angle =0]{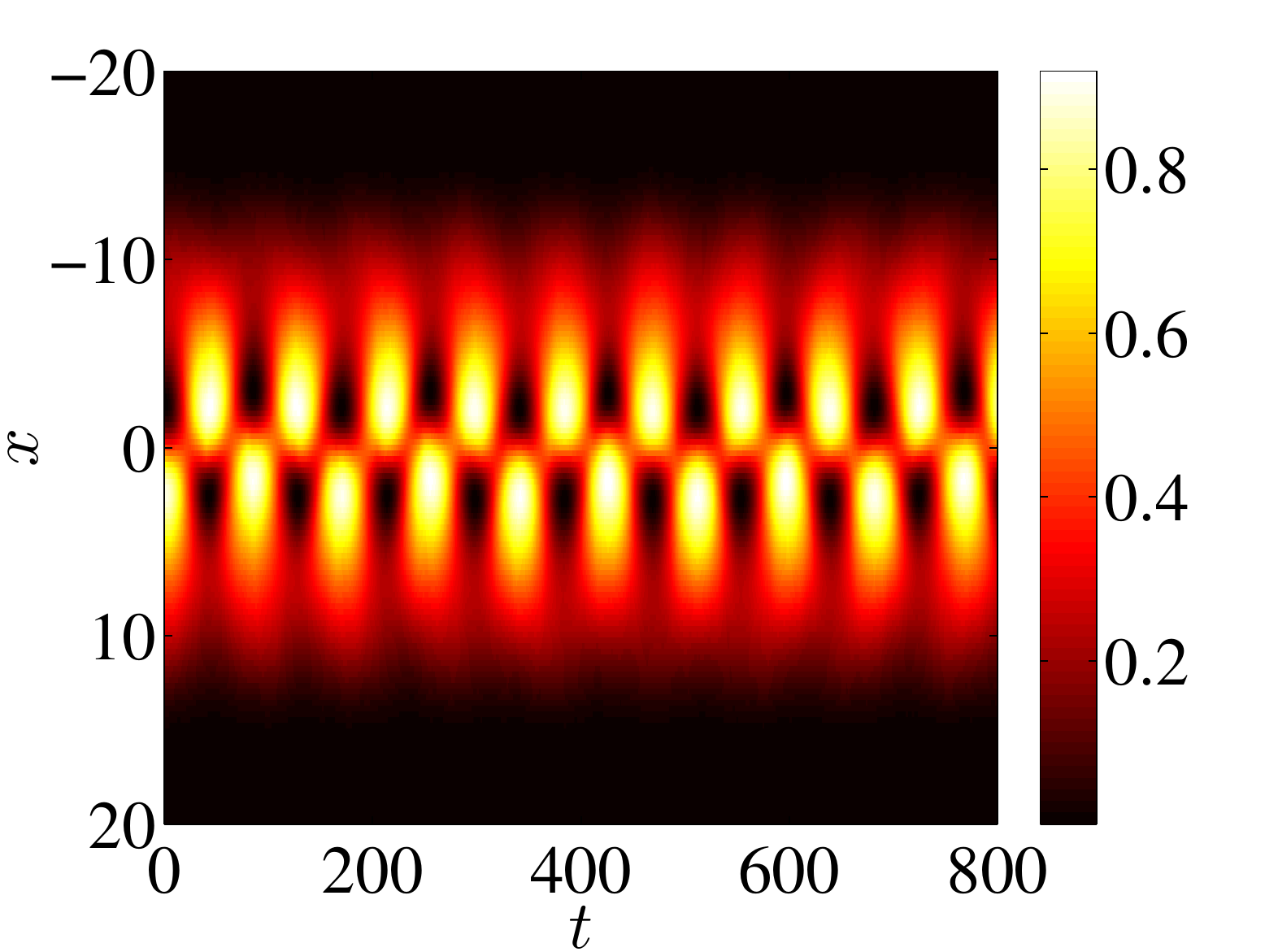}
\label{fig2l}
}
}
\end{center}
\par
\vspace{-0.7cm}
\caption{
(Color online) Same as Fig.~\ref{fig1} but for unequal
interaction coefficients, with $g_{11}=1.03$, $g_{12}=1$ and
$g_{22}=0.97$.
}
\label{fig2}
\end{figure}

Our 1D results are summarized in Figs.~\ref{fig1} and \ref{fig2}.
In particular, Fig.~\ref{fig1} corresponds to the case of equal 
interaction coefficients, while results obtained using unequal 
interaction coefficients are presented in Fig.~\ref{fig2}. The 
DB solitons corresponding to the fundamental ingredients for our
study are depicted in the left columns of Figs.~\ref{fig1} and~\ref{fig2}
with dashed-dotted black and blue lines, respectively, while their
rotated siblings by $\pi/8$ and $\pi/4$ are presented with solid
black and blue lines therein. It is worth pointing out that the 
stability trait of the original, i.e., unrotated DB soliton states
employed here (with and without a trap) has been extensively studied; 
for a recent example see, e.g., Ref.~\cite{dbs_coupled_Manakov} and 
references therein. This way, the underlying unrotated states for 
values of the chemical potentials of $\mu_{-}=1$ and $\mu_{+}=0.9$ 
are stable.

Having identified the states of interest, we now turn our discussion
to the dynamical evolution of the ($SO(2)$) rotated variants of DB 
solitons, namely the DD states, and monitor their oscillatory development. 
Specifically, the middle and right panels of Figs.~\ref{fig1} and~\ref{fig2}
present the spatio-temporal evolution of the densities $|\Phi_{-}(x,t)|^2$
and $|\Phi_{+}(x,t)|^{2}$, respectively (hereafter, for simplicity, we 
omit primes in the rotated fields). From these panels, the development of
the well-known beating DD soliton~\cite{pe4,pe5}, showcasing a breathing
oscillation of the corresponding individual densities, is clearly evident. 
Furthermore, the oscillation persists over a wide time interval of integration
forward in time (note the range of the $t$-axis in these panels), while these
findings indicate the robustness of such states which is also expected
since they were also observed in experiments~\cite{pe4,pe5}. 

Let us highlight some differences between the integrable (i.e., equal
interaction coefficients) and the non-integrable cases, that are apparent
not only in the 1D setting, but in the 2D as well as 3D settings
which will be discussed next. It can be discerned from panels (b) and
(c), as well as (e) and (f) of Figs.~\ref{fig1} and~\ref{fig2}, 
where the trap is absent, that robust beating solitons form and
oscillate with a fixed period of oscillation [cf. Eq.~(\ref{period})]. 
However, as soon as we
depart from the integrable case, the period of oscillations is affected
due to the fact that the $SU(2)$-invariance is broken away from this limit.
In particular, it is evident from these panels of Fig.~\ref{fig2} that
the period increases. We note in passing that small amount of radiation
is observed as well (see, e.g., panels (e) and (f) in Fig.~\ref{fig2}), 
which affects the period of oscillations. Similar findings are reported
for the case with a trap as depicted in panels (h) and (i), and (k) and
(l) of Figs.~\ref{fig1} and~\ref{fig2}. In all of these cases,
the excitation persists. While the presence of the trap does not seem
to dramatically affect its (internal) dynamics, nevertheless, 
when departing from the equal interaction case, it does appear to 
affect its details. Notice, in particular, the vibration frequency [%
cf. the beating differences between the first and third, second and
fourth row of panels in Fig.~\ref{fig2}].

\subsection{Vortex--Vortex
structures in 2D}

\begin{figure}[t]
\begin{center}
\vspace{-0.1cm}
\mbox{\hspace{-0.1cm}
\subfigure[][]{\hspace{-0.3cm}
\includegraphics[height=.22\textheight, angle =0]{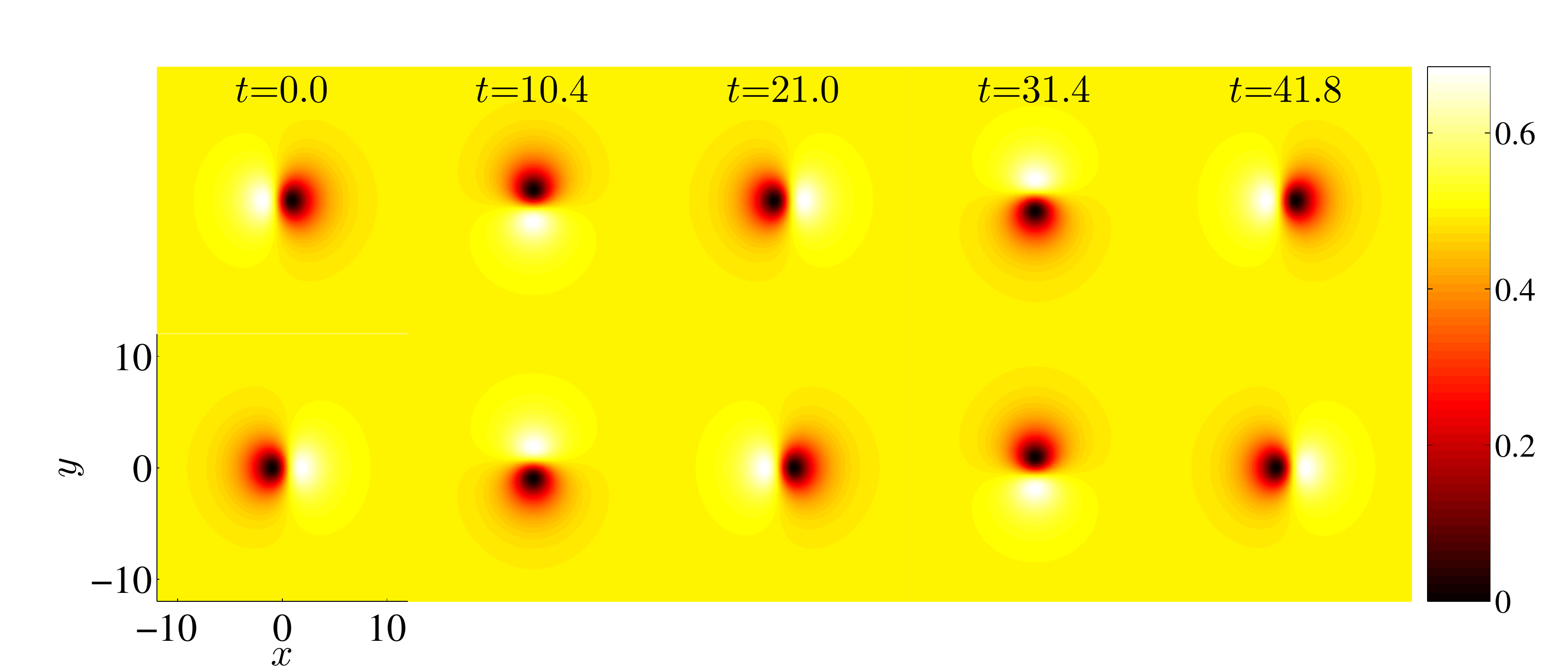}
\label{fig3a}
}
}
\mbox{\hspace{-0.1cm}
\subfigure[][]{\hspace{-0.3cm}
\includegraphics[height=.20\textheight, angle =0]{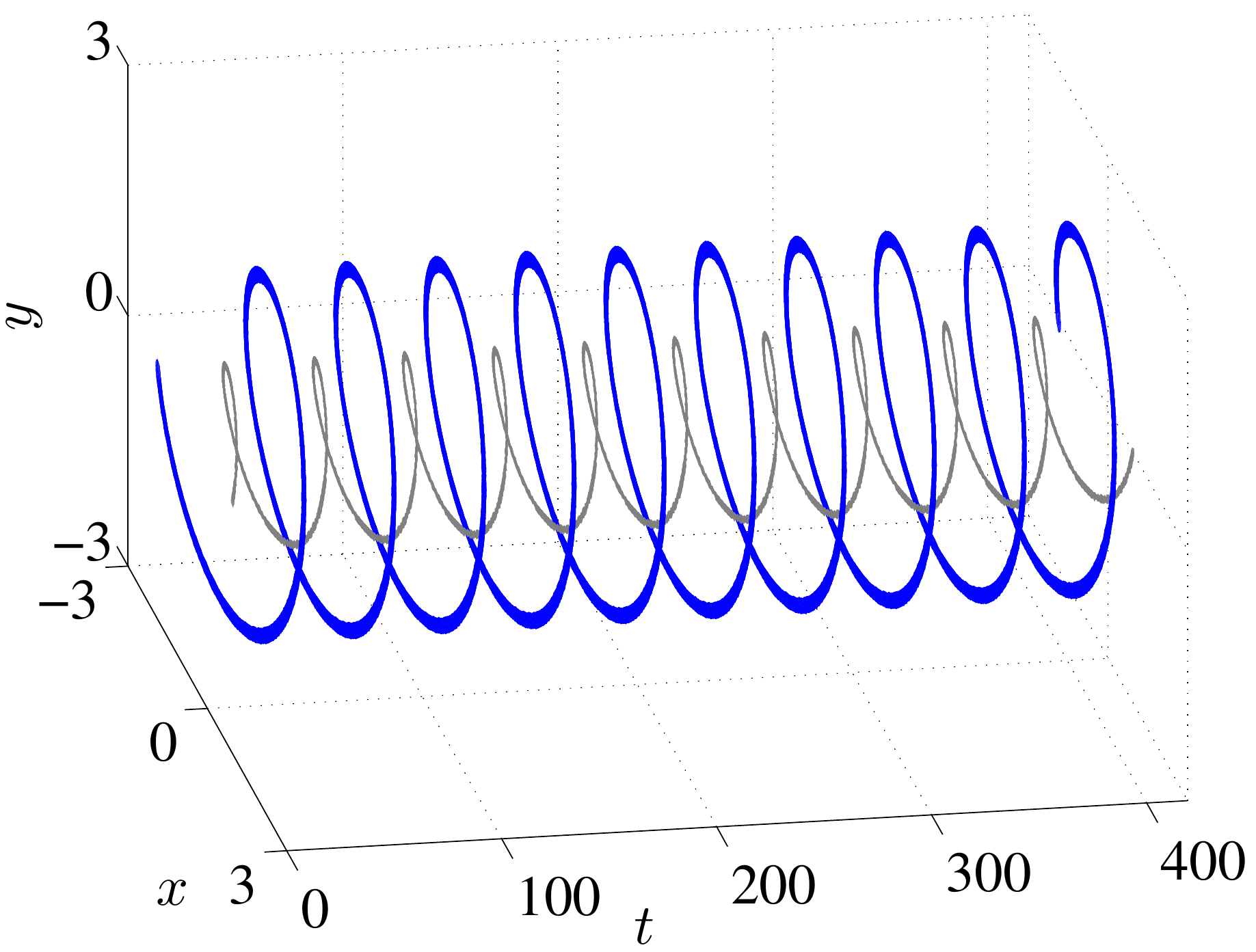}
\label{fig3b}
}
\subfigure[][]{\hspace{-0.3cm}
\includegraphics[height=.20\textheight, angle =0]{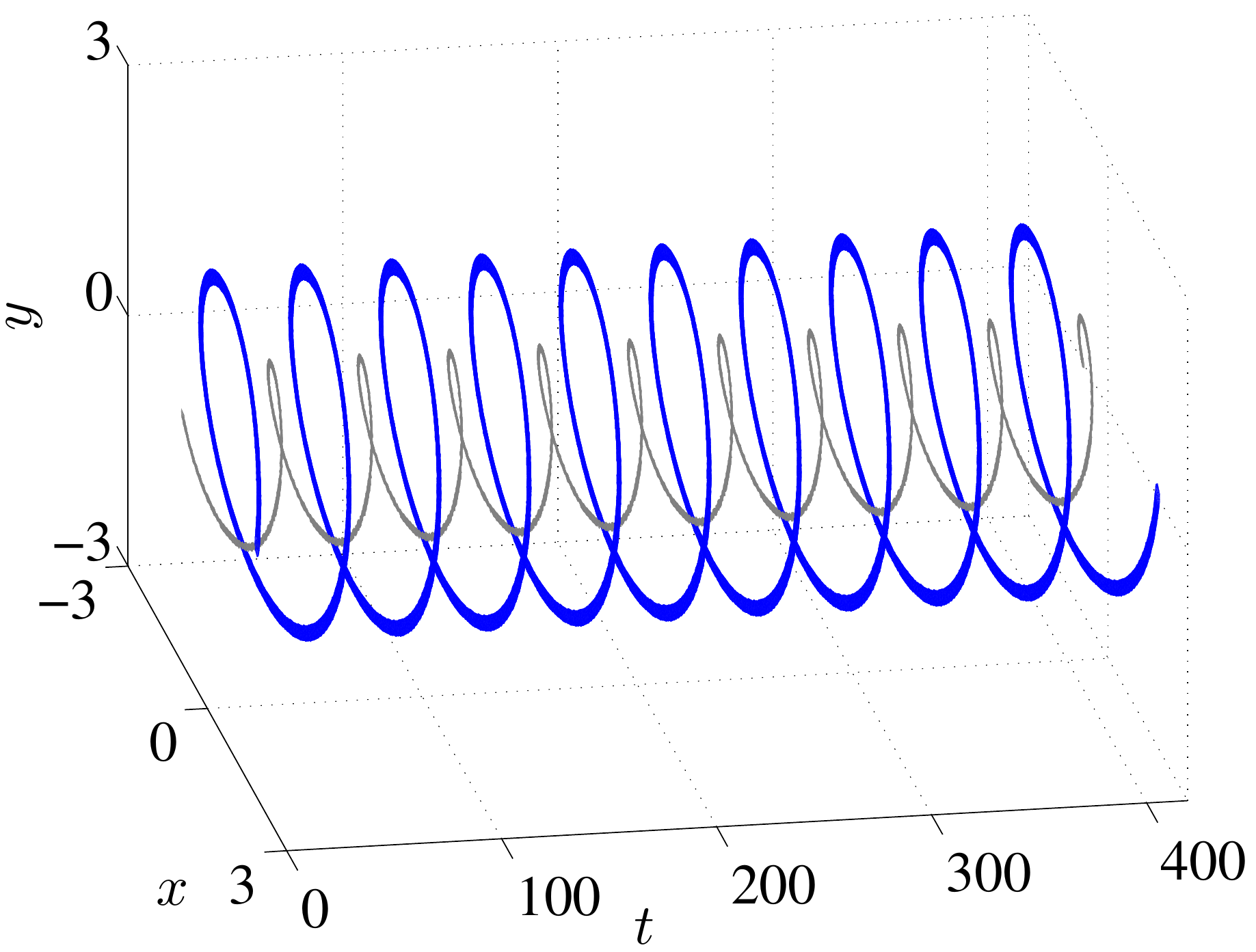}
\label{fig3c}
}
}
\end{center}
\par
\vspace{-0.7cm}
\caption{
(Color online) Summary of results under the action of the $SO(2)$
rotation by $\pi/4$ corresponding to the homogeneous case with
equal interaction coefficients and values of the chemical potentials
of $\mu_{-}=1$ and $\mu_{+}=0.85$. \textit{Top row}: Snapshots of
densities $|\Phi_{-}(x,y,t)|^{2}$ (top panels) and $|\Phi _{+}(x,y,t)|^{2}$
(bottom panels) at different instants of time. \textit{Bottom row}:
Isosurfaces of the spatiotemporal evolution of the densities
$|\Phi_{-}(x,y,t)|^{2}$ and $|\Phi _{+}(x,y,t)|^{2}$ presented in
panels (b) and (c), respectively. Each isosurface depicted by
blue and gray color corresponds to a value of $0.001\times \max(|\Phi_{-}(x,y,t)|^{2})$
and $0.999\times\max(|\Phi_{+}(x,y,t)|^{2})$, respectively.
}
\label{fig3}
\end{figure}

\begin{figure}[t]
\begin{center}
\vspace{-0.1cm}
\mbox{\hspace{-0.1cm}
\subfigure[][]{\hspace{-0.3cm}
\includegraphics[height=.22\textheight, angle =0]{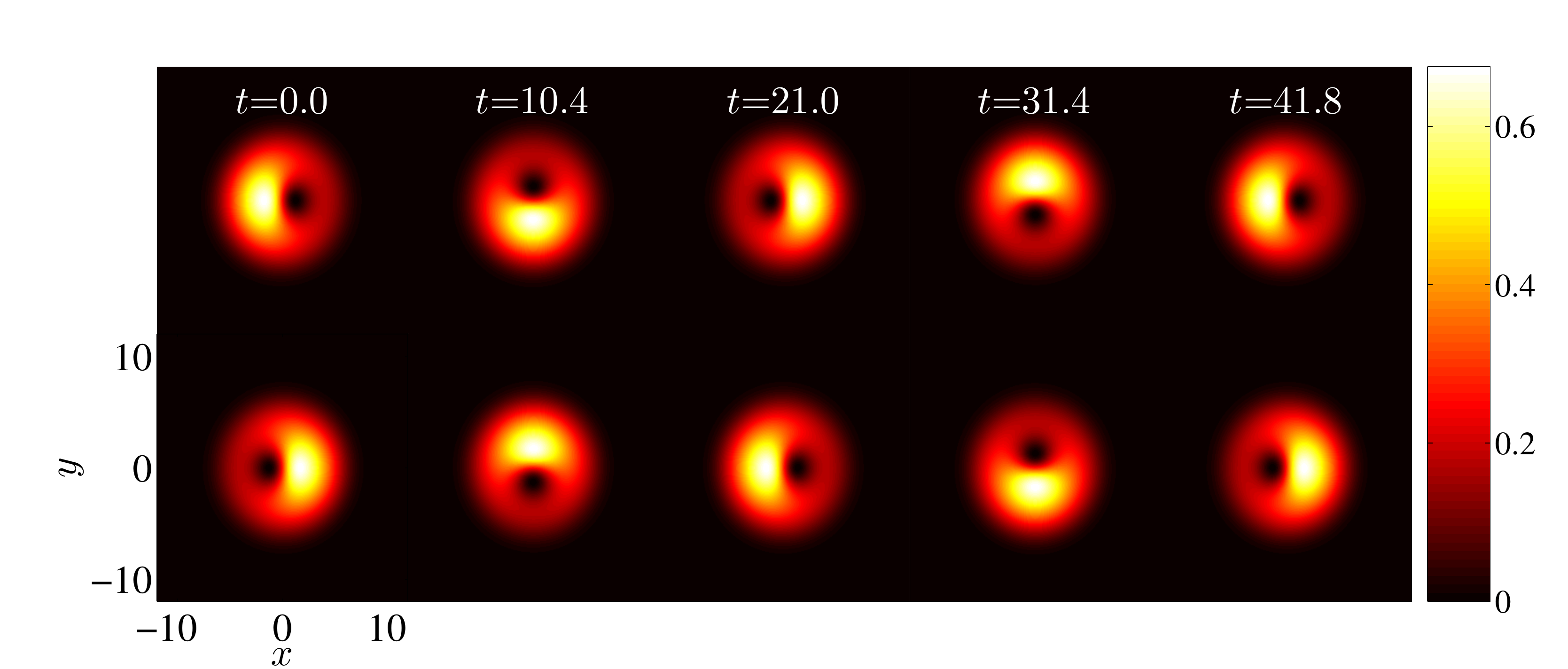}
\label{fig4a}
}
}
\mbox{\hspace{-0.1cm}
\subfigure[][]{\hspace{-0.3cm}
\includegraphics[height=.20\textheight, angle =0]{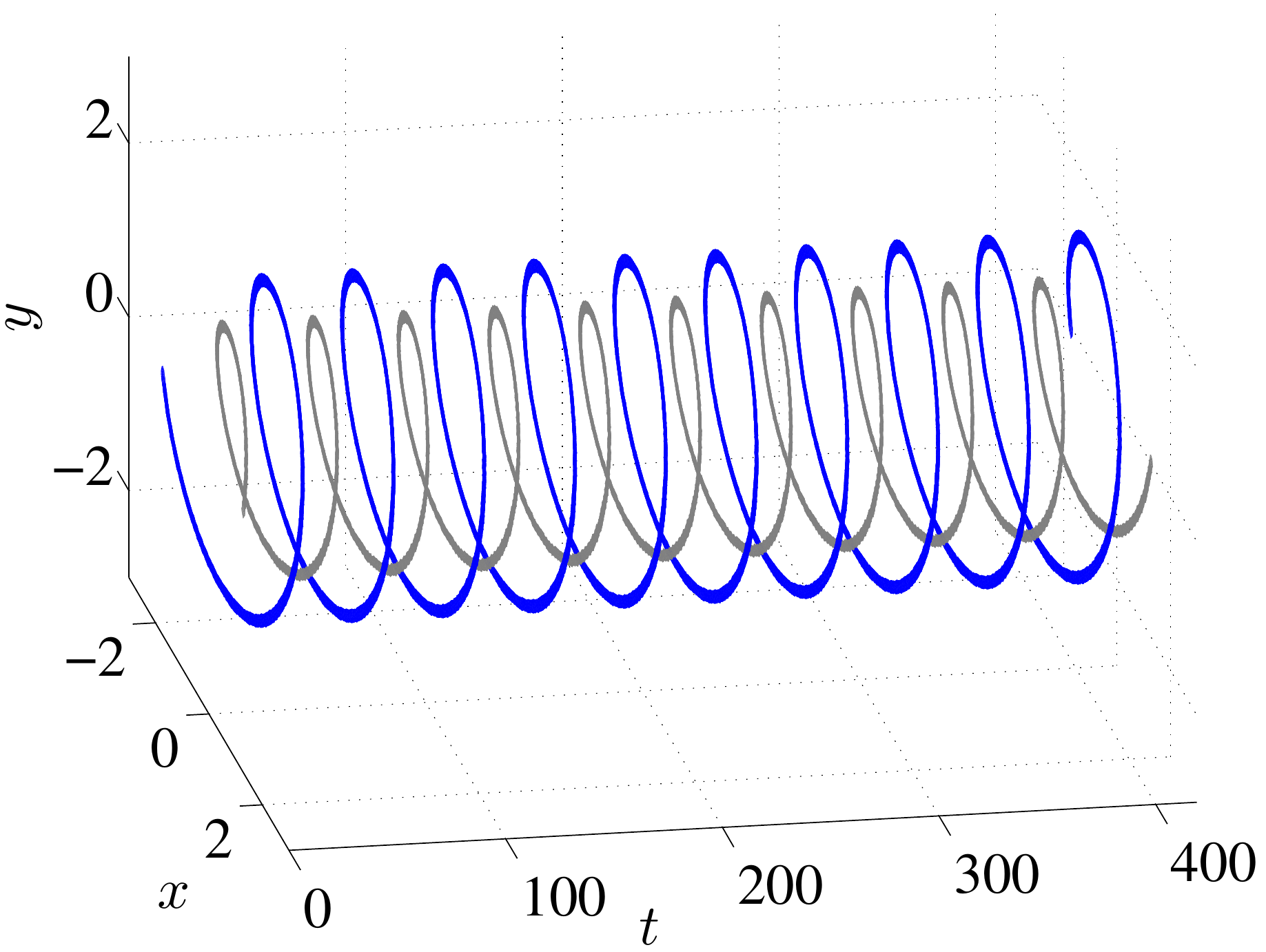}
\label{fig4b}
}
\subfigure[][]{\hspace{-0.3cm}
\includegraphics[height=.20\textheight, angle =0]{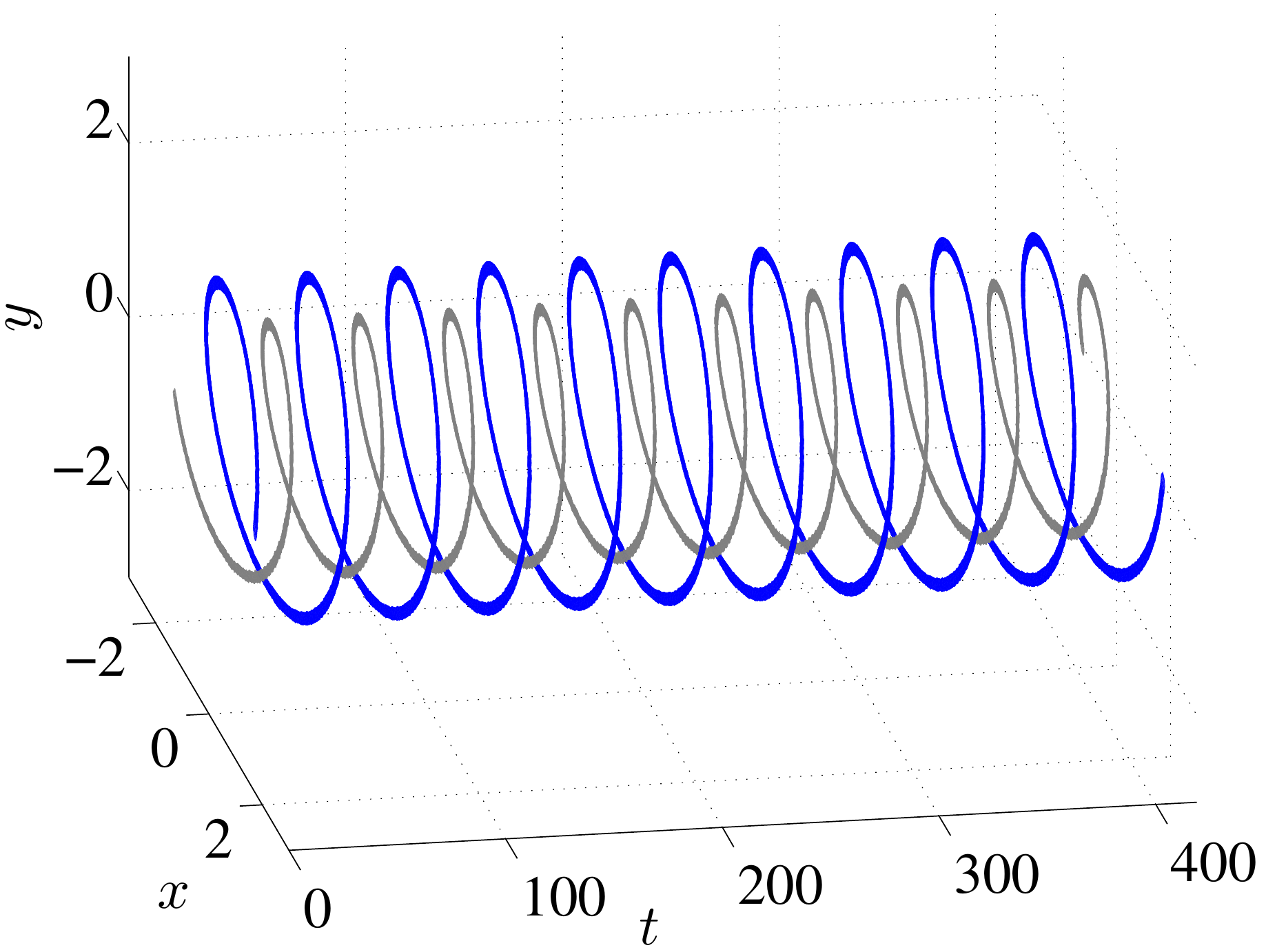}
\label{fig4c}
}
}
\mbox{\hspace{-0.1cm}
\subfigure[][]{\hspace{-0.3cm}
\includegraphics[height=.15\textheight, angle =0]{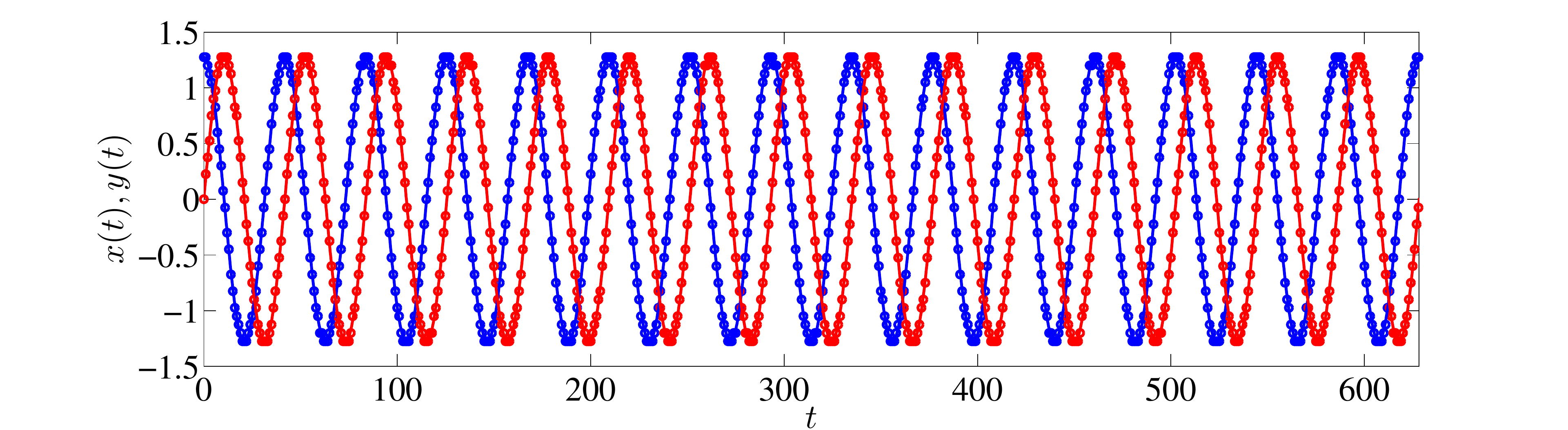}
\label{fig4d}
}
}
\end{center}
\par
\vspace{-0.7cm}
\caption{
(Color online) Same as Fig.~\ref{fig3} but under the presence of
harmonic confinement with $\Omega=0.2$. \textit{Top row}: Snapshots
of densities $|\Phi_{-}(x,y,t)|^{2}$ (top panels) and $|\Phi _{+}(x,y,t)|^{2}$
(bottom panels) at different instants of time. \textit{Middle row}:
Isosurfaces of the spatiotemporal evolution of the densities
$|\Phi_{-}(x,y,t)|^{2}$ and $|\Phi _{+}(x,y,t)|^{2}$ presented
in panels (b) and (c), respectively. Each isosurface depicted by
blue and gray color corresponds to a value of $0.001\times \max(|\Phi_{-}(x,y,t)|^{2})$
and $0.999\times\max(|\Phi_{+}(x,y,t)|^{2})$, respectively.
\textit{Bottom row}: The location of the vortex $(x,y)$ in the
first component as a function of time where its abscissa and
ordinate are depicted with blue and red circles, respectively.
The solid blue and red lines correspond to the theoretical prediction.
}
\label{fig4}
\end{figure}

\begin{figure}[t]
\begin{center}
\vspace{-0.1cm}
\mbox{\hspace{-0.1cm}
\subfigure[][]{\hspace{-0.3cm}
\includegraphics[height=.22\textheight, angle =0]{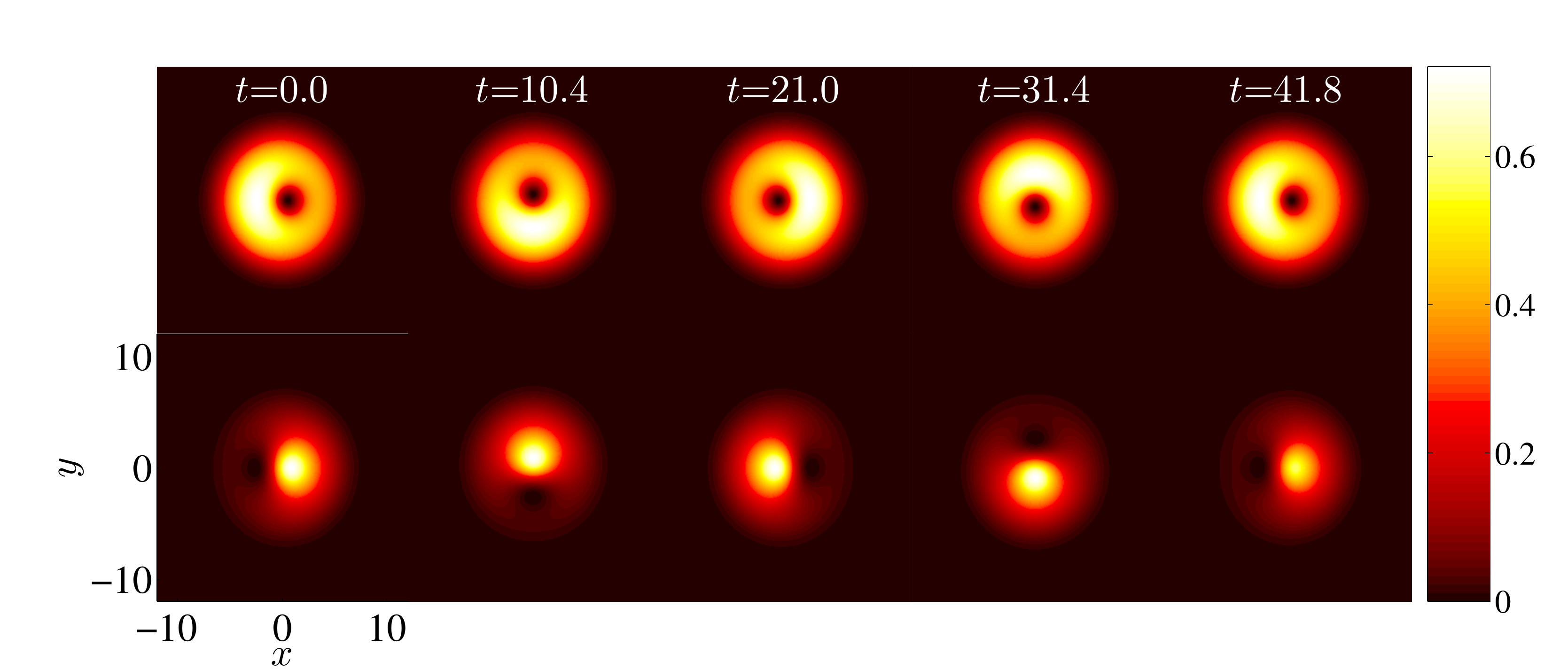}
\label{fig5a}
}
}
\mbox{\hspace{-0.1cm}
\subfigure[][]{\hspace{-0.3cm}
\includegraphics[height=.20\textheight, angle =0]{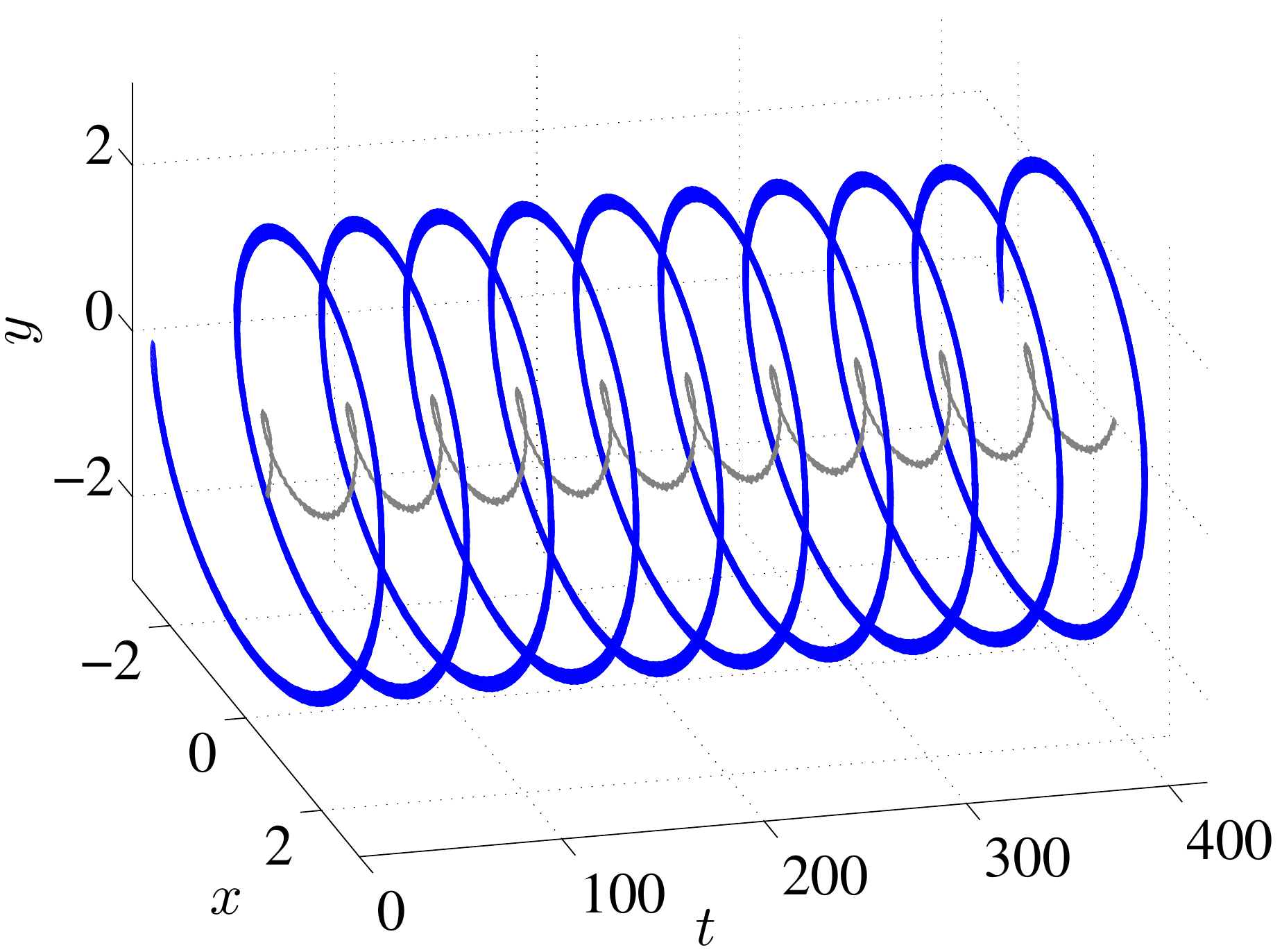}
\label{fig5b}
}
\subfigure[][]{\hspace{-0.3cm}
\includegraphics[height=.20\textheight, angle =0]{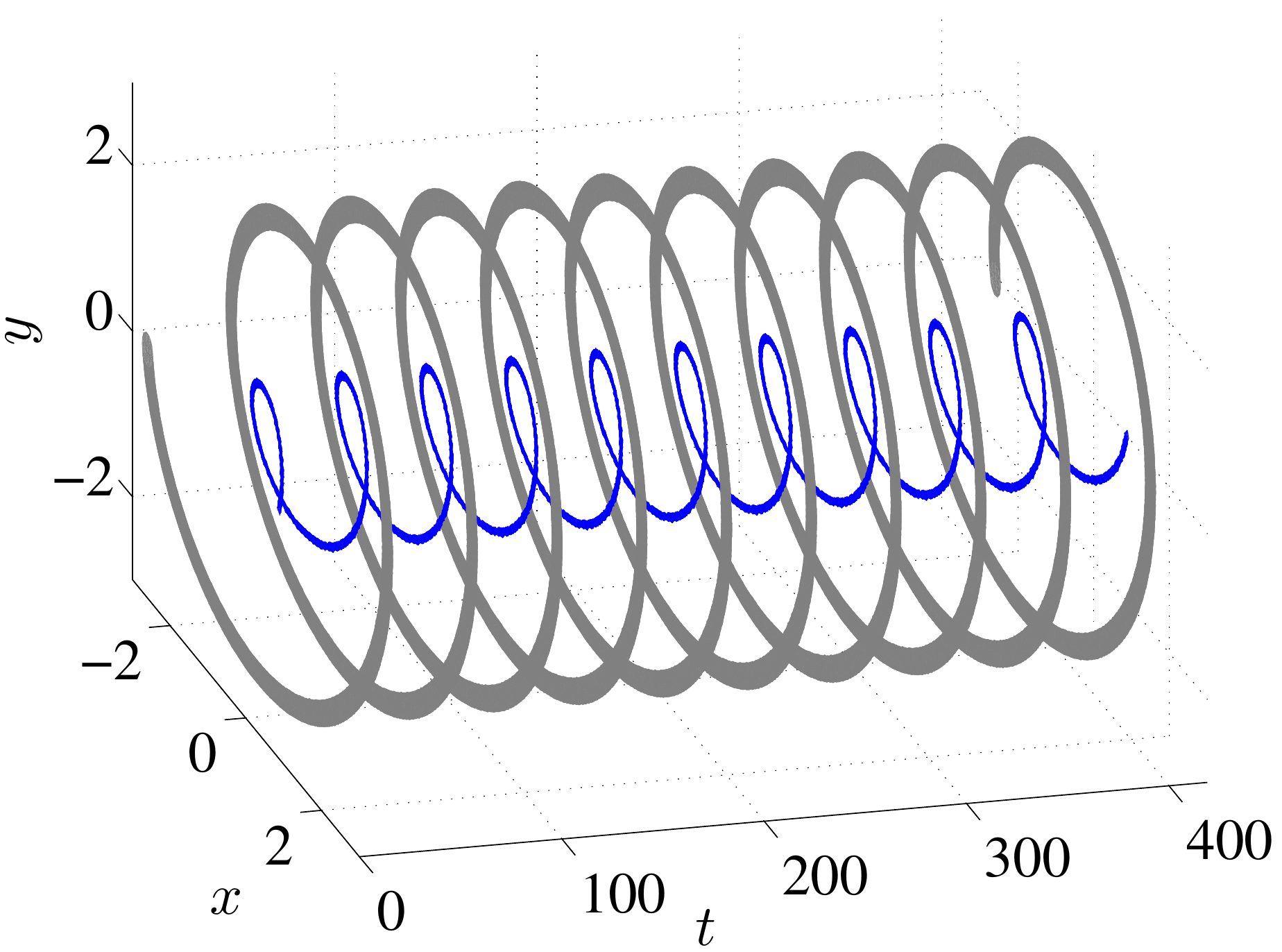}
\label{fig5c}
}
}
\end{center}
\par
\vspace{-0.7cm}
\caption{
(Color online) Same as Fig.~\ref{fig4} but for $\delta=\pi/8$.
\textit{Top row}: Snapshots of densities $|\Phi_{-}(x,y,t)|^{2}$
(top panels) and $|\Phi _{+}(x,y,t)|^{2}$ (bottom panels) at
different instants of time. \textit{Bottom row}: Isosurfaces of
the spatiotemporal evolution of the densities $|\Phi_{-}(x,y,t)|^{2}$
and $|\Phi _{+}(x,y,t)|^{2}$ presented in panels (b) and (c),
respectively. Similarly, each isosurface depicted by blue and gray
color corresponds to a value of $0.001\times \max(|\Phi_{-}(x,y,t)|^{2})$
and $0.999\times\max(|\Phi_{+}(x,y,t)|^{2})$, respectively.
}
\label{fig5}
\end{figure}

\begin{figure}[t]
\begin{center}
\vspace{-0.1cm}
\includegraphics[height=.22\textheight, angle =0]{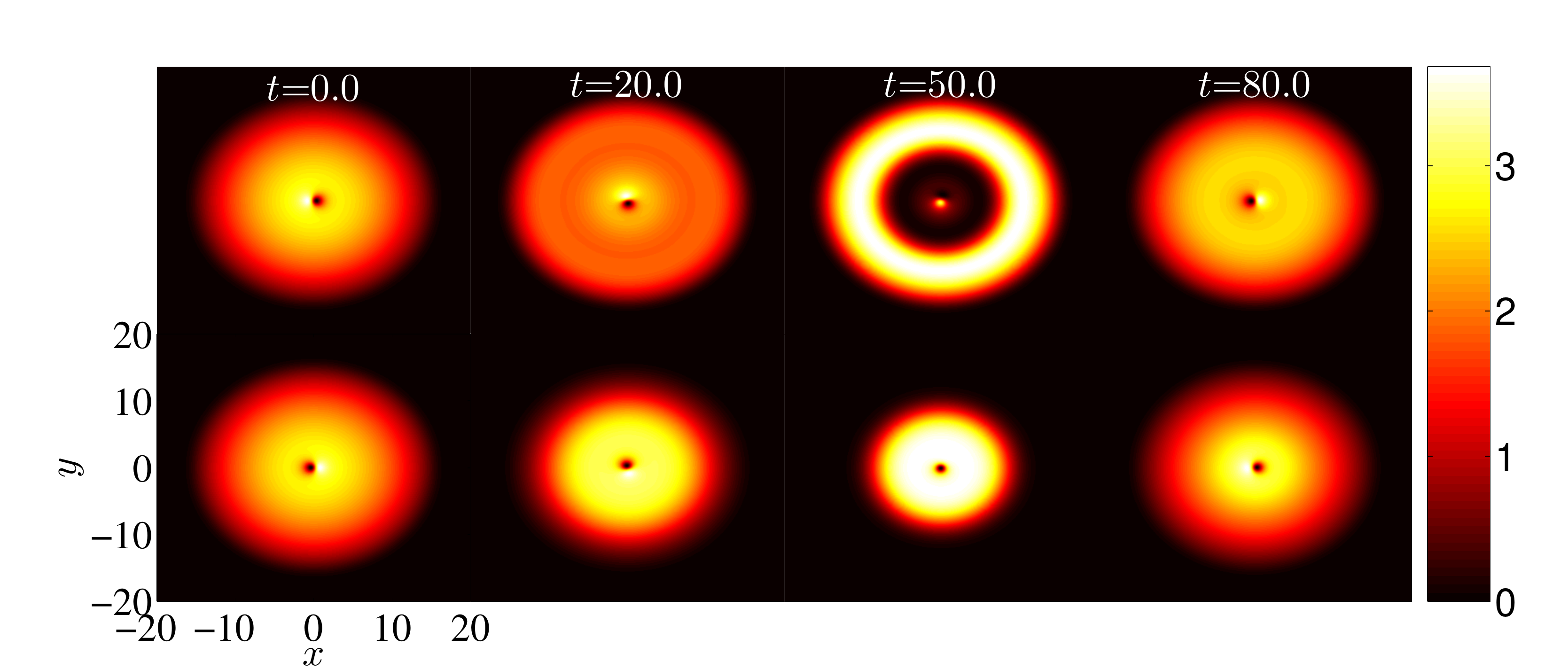}
\end{center}
\par
\vspace{-0.7cm}
\caption{
(Color online) Same as Fig.~\ref{fig4} but for unequal
interaction coefficients. Snapshots of densities $|\Phi_{-}(x,y,t)|^{2}$
(top panels) and $|\Phi _{+}(x,y,t)|^{2}$ (bottom panels) at different
instants of time. Here, the phase separation in this immiscible
regime affects the density by resulting in target patterns (previously
observed also in experiments; see the text) in the dynamics. See 
Ref.~\cite{vv_movie} for a more complete movie of the dynamics.
}
\label{fig6}
\end{figure}

In this section, we take a step further and discuss rotated
vortex-bright soliton complexes in 2D, by considering specific cases
spanning various possibilities. Figures~\ref{fig3}, \ref{fig4}
and \ref{fig5} depict examples of initially rotated -- so as
to produce the vortex--vortex (VV) state -- and dynamically
evolved vortex-bright solitons with equal interaction coefficients and
values of the chemical potentials of $\mu_{-}=1$ and $\mu_{+}=0.85$.
Figure~\ref{fig6}, corresponding to $\mu_{-}=5.2$ and $\mu_{+}=4.2$,  
highlights the effect of unequal interaction coefficients.
Furthermore, Figs.~\ref{fig3} and~\ref{fig4} correspond to a rotation
by $\pi/4$ of the original vortex-bright soliton complex, 
in the absence
and presence of a trap (with $\Omega=0.2$), respectively. On the same
footing, Fig.~\ref{fig5} corresponds to a rotation by $\pi/8$ of the
original state whereas Fig.~\ref{fig6} for 
$\mu_{-}=5.2$ and $\mu_{+}=4.2$ involves rotation by $\pi/4$.
Both of the latter examples are in the presence 
of a trap. Snapshots of the densities $|\Phi_{-}(x,y,t)|^2$ and
$|\Phi_{+}(x,y,t)|^{2}$ at different instants of time $t$ are depicted
in the top and bottom rows, respectively, of Figs.~\ref{fig3a}, \ref{fig4a},
and \ref{fig5a}, as well as in Fig.~\ref{fig6}. Our study is complemented by
demonstrating isocontours of the individual densities of the vortex and 
bright soliton of each component in panels (b) and (c) of Figs.~\ref{fig3},
\ref{fig4} and \ref{fig5} with gray and blue colors, respectively.

As has been illustrated in the recent study of~\cite{vbs_coupled}
(even in the absence of a trap), but also in earlier works
in the presence of a trap~\cite{kodyprl}, the vortex-bright
state is generally stable. As mentioned in Section~II, its 
rotated VV counterpart inherits these traits. Furthermore, 
the internal period $T$ of vibration of the VV state in the
equal interactions coefficients case is given by Eq.~(\ref{period}), 
as shown in the previous section. In our numerical results 
the period calculated numerically follows 
this analytical prediction, a feature that we have used as a 
benchmark of our numerical method \cite{footnote}. Once again, 
the presence of the trap does not appear to significantly affect 
the motion of the vortices in the case of equal interaction coefficients:
the vortex constituents of the VV state in each component continue
to blithely orbit around each other both in the presence and in 
the absence of the trap. 

Specifically, snapshots of the densities are presented in panels (a)
of Figs.~\ref{fig3}, \ref{fig4} and \ref{fig5} at each $t$ which is
equal to one quarter of the period, i.e., $t=0$, $t=T/4$, $t=T/2$,
$t=3T/4$ and $t=T$ (with $T\approx 41.88$ in these examples). This
way, the vortex--vortex complex performs a circular motion as time
evolves (see, the insets therein) and returns to its original position
at $t=T$ (see, the last column of panels (a)). Furthermore, the 
oscillations of the vortex-vortex complexes are persistent as our 
long-time dynamics reveal in panels (b) and (c) of Figs.~\ref{fig3},
\ref{fig4} and \ref{fig5} (see, the range of $t$ axes therein) suggesting
that the underlying states are indeed robust.

Arguably more intriguing, however, appears to be the case of unequal 
interaction coefficients. In this case, 
and in the presence of the trap, the results are illustrated
in Fig.~\ref{fig6}; see, also,~\cite{vv_movie} for a complete
movie of the dynamics in this case. Although the initial vortex--bright 
soliton is stable (in the realm of linear stability analysis), its 
vortex--vortex sibling appears to undergo modifications of its density
profile. At first, the vortex--vortex complex follows a circular motion,
where the period increases compared to the analytical prediction of 
Eq.~(\ref{period}), due to the unequal interaction coefficients. 
In analogy to the 1D setting, this is expected based on 
the fact that the $SU(2)$-invariance is broken. Then, the 
configuration starts
changing in shape (see the panels in the second column of the Figure) 
leading at $t=50$ the bright soliton in the second component to disappear, 
while in the first component the vortex structure cannot be straightforwardly
discerned in the density. However, the complex in the second component 
regains (qualitatively) its structural form back around $t=80$, leading to 
the recurrence of the VV state (in a rotated form). It is evident in the
snapshots (especially of the second and third column of Fig.~\ref{fig6}) 
that the dynamics features phase separation phenomena analyzed in detail,
e.g., in the experimental (and computational) analysis 
of Ref.~\cite{originalhall};
see also~\cite{refhall}. Indeed, while the rotation of the VV pattern persists
(or, at least, recurs), the overall density pattern develops the target 
patterns analyzed in (the planar projections associated with) Ref.~\cite{originalhall};
see also \cite{siambook} and references therein. Our conclusion in that 
connection is that the robustness of the rotational state is, at least in
part, affected by the location of the relevant interaction coefficients 
with respect to the miscibility/immiscibility transition -- associated
with crossing the critical point $D=0$ of the immiscibility parameter 
$D \equiv g_{11} g_{22}- g_{12}^2$~\cite{siambook}.


\subsection{VL--VL and VR--VR solitons in 3D}

\begin{figure}[tb]
\begin{center}
\includegraphics[scale=0.11]{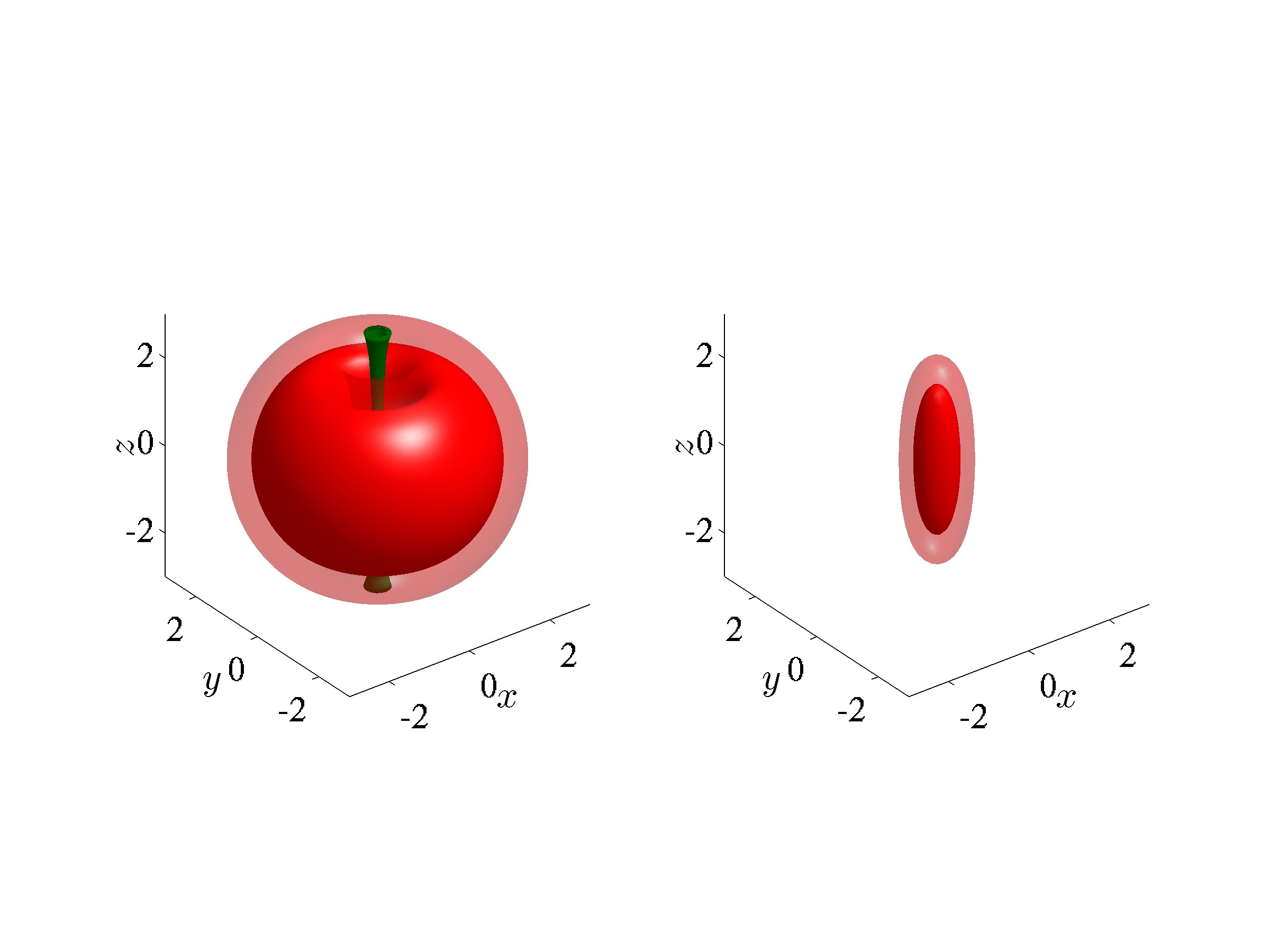}
\caption{(Color online)
The density isocontour plots of a stable stationary vortex-line-bright
soliton at $\mu_1=7$ and $\mu_2=6.2$ in an isotropic trap with $\Omega=1$.
The core of the line is highlighted in green (dark) contours.
}
\label{VLB5}
\end{center}
\end{figure}

Finally, we study the effect of $SO(2)$ rotations to construct 
vorticity-bearing vector  structures in 3D. In particular, we focus on the
cases of the VL-bright soliton and the VR-bright soliton. As in 
1D and 2D, we first identify the stationary states, study their 
stability traits and subsequently rotate the corresponding states.
Then, we monitor the dynamics of these states by advancing the 
NLS system forward in time. A stationary VL--bright soliton state 
is shown is Fig.~\ref{VLB5}. We have checked that the state at the
studied parameters is stable using spectral stability analysis methods
analogous to those utilized in Ref.~\cite{dsss_ww}, as well as direct 
dynamical integration up to $t=100$. 

\begin{figure}[tb]
\begin{center}
\includegraphics[scale=0.11]{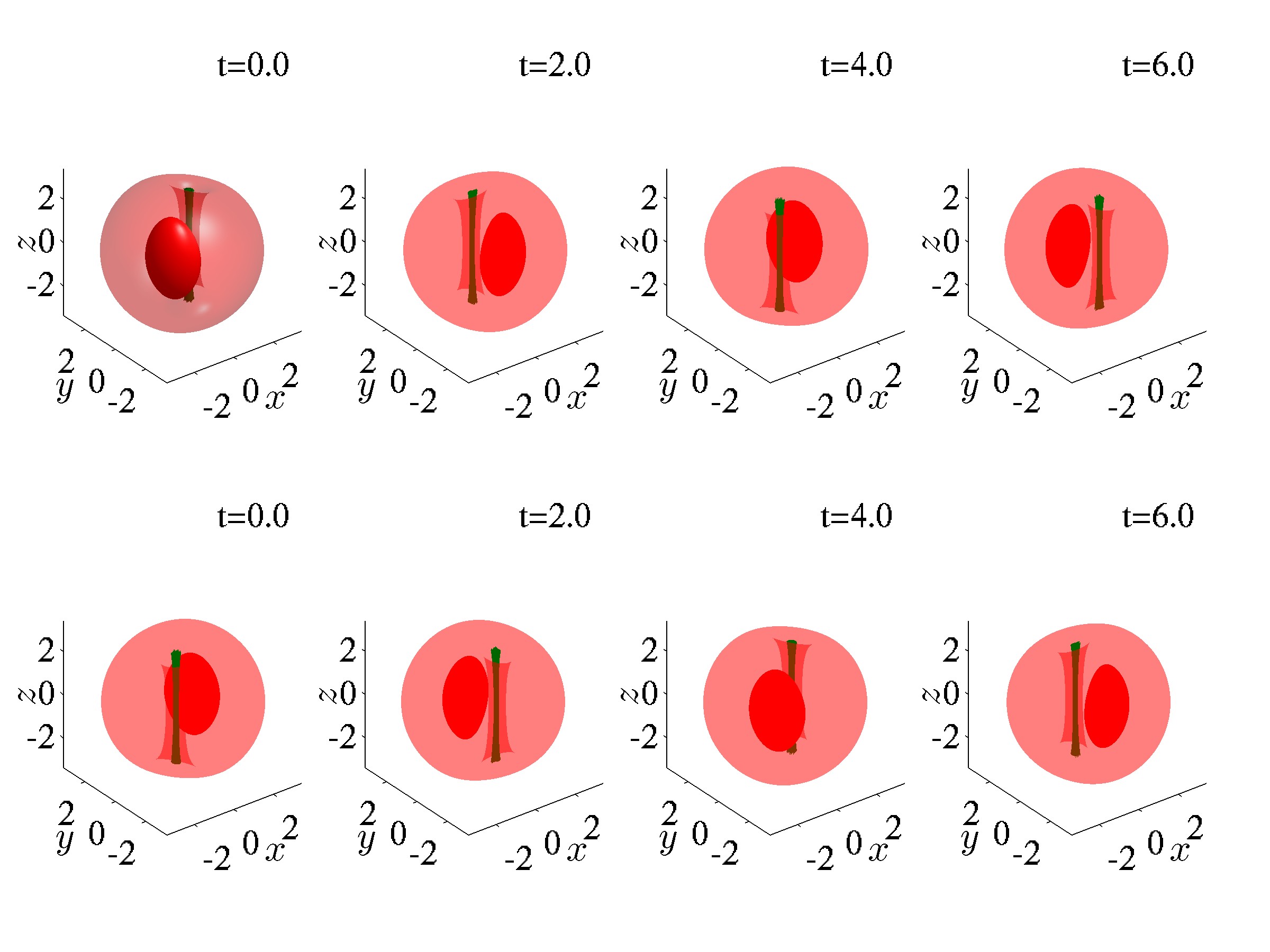}
\caption{(Color online)
Robust VL--VL oscillations transformed from the VL-bright soliton state
shown in Fig.~\ref{VLB5}. Typical states are shown for one period, with
the top panel for one component and the bottom panel for the other component.
See Ref.~\cite{VLVL1_movie} for a more complete movie of the dynamics.
}
\label{VLBD}
\end{center}
\end{figure}

Subsequently, we perform the $SO(2)$ rotation with $\delta=\pi/4$, 
and the  VL--bright soliton state morphs into a VL-VL solitary
wave. Dynamical evolution shows that the two vortex lines perform a
rotational motion around each other in the trap. Some typical intermediate
stages within a period are shown in Fig.~\ref{VLBD}. See Ref.~\cite{VLVL1_movie}
for a more complete movie of the dynamics. We have also verified that similar
robust dynamics also hold for $\delta=\pi/8$. Hence, such VL--VL states
are natural candidates for observation in the dynamics of the system 
-- although, of course, it does not escape us that unequal interaction
coefficients may again impose density modulations via phase-separation
phenomena; we comment on this further below.

\begin{figure}[tb]
\begin{center}
\includegraphics[scale=0.11]{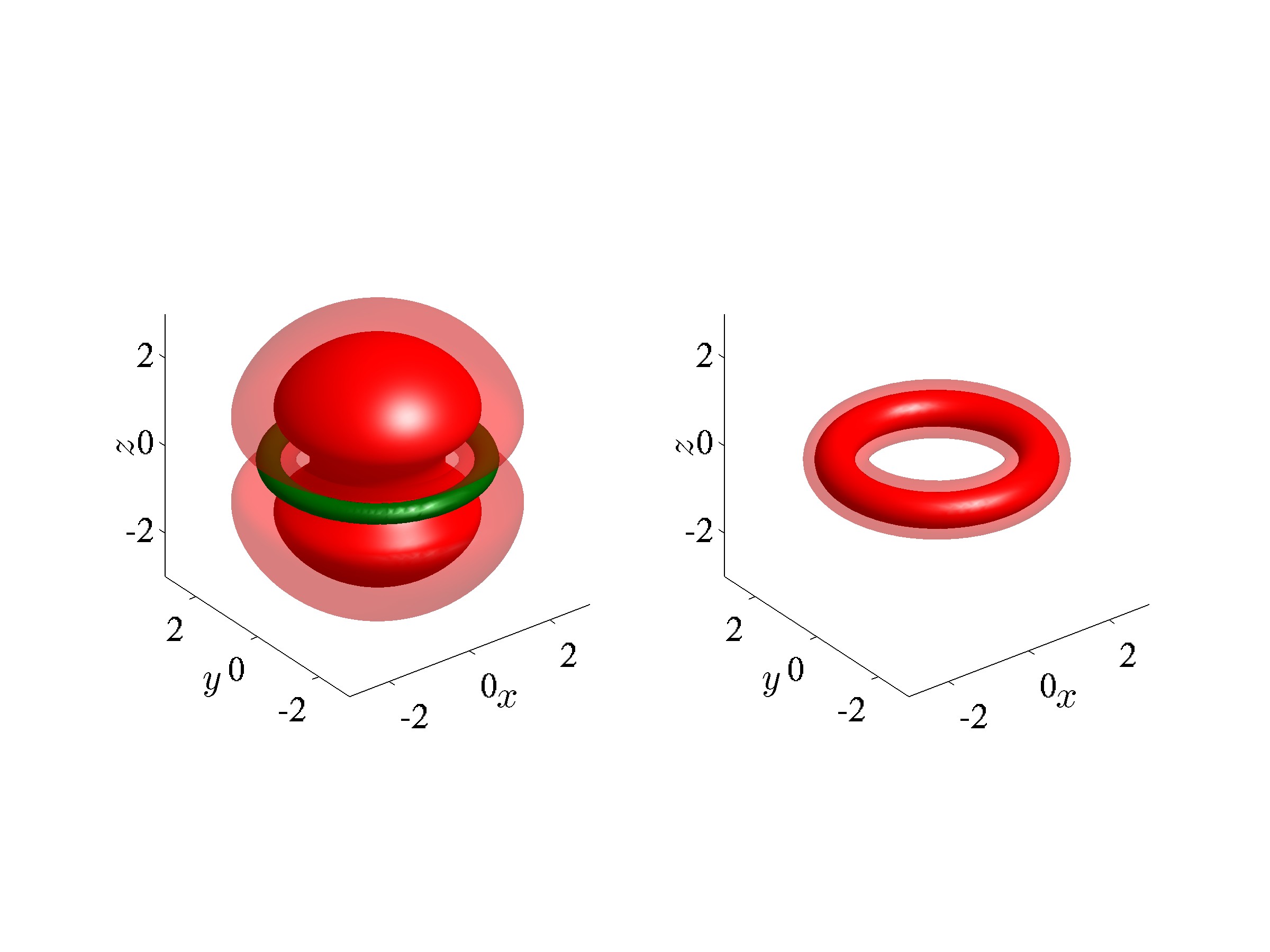}
\caption{(Color online)
The density isocontour plots of a stable stationary vortex-ring-bright soliton
at $\mu_1=9$ and $\mu_2=7.6$ in an isotropic trap with $\Omega=1$. The core
of the ring is highlighted in green (dark) contours.
}
\label{VRB5}
\end{center}
\end{figure}

\begin{figure}[tb]
\begin{center}
\includegraphics[scale=0.11]{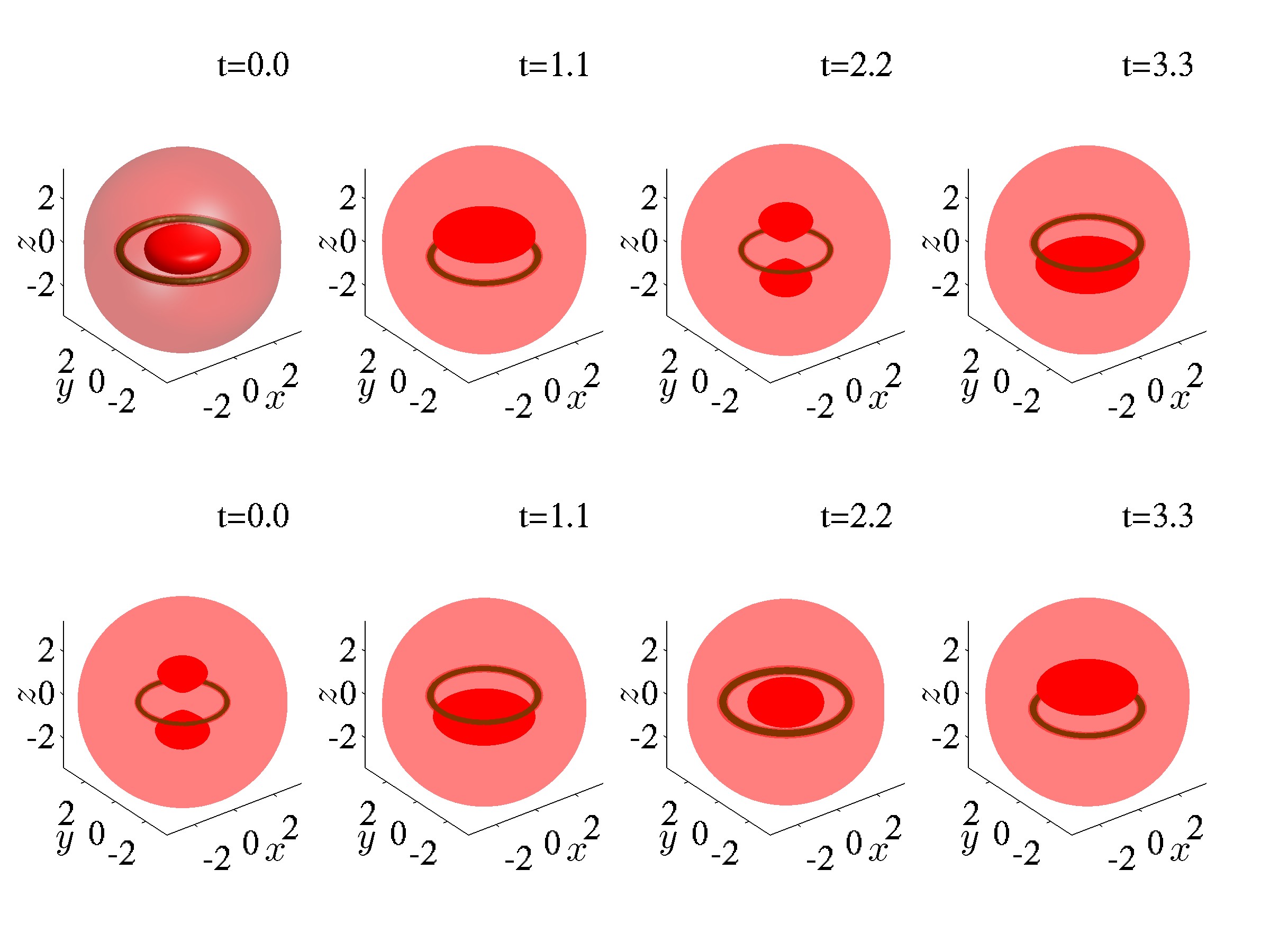}
\caption{(Color online)
Robust VR--VR oscillations transformed from the VR-bright soliton state shown in
Fig.~\ref{VRB5}. Typical states are shown for one period, with the top panel for
one component and the bottom panel for the other component. See Ref.~\cite{VRVR1_movie}
for a more complete movie of the dynamics.
}
\label{VRBD}
\end{center}
\end{figure}

Now we discuss the VR--bright soliton. Similarly, a stable stationary state
of the VL--bright soliton is shown in Fig.~\ref{VRB5}, and is converged upon
fixed point iteration. Dynamics in the case of $\delta=\pi/4$ are shown in 
Fig.~\ref{VRBD}, and robust oscillations also hold for the case $\delta=\pi/8$. 
Here, the vortex rings are involved in an intriguing ``dance'' routine, 
where they vibrate between pairs of inner-outer, then top-bottom, then outer-inner,
and finally bottom-top (for the two species), before the cycle restarts, as 
is 
illustrated in the figure. A more detailed perspective of the relevant
choreography  is given in the movie of 
Ref.~\cite{VRVR1_movie}.

It is important to remind the reader here that these results were
obtained with a fairly confining isotropic trap of strength 
$\Omega=1$. We have explored the dynamics observed in the
case of $g_{ij} \neq 1$ for phase separation and while we did see
signatures of the latter, these were found to be quite weak in 
this setting (relatively to the 2D case discussion presented above). 
This is in line with earlier observations, see 
e.g.,~\cite{rafael}, indicating theoretically and computationally
that the phase separation transition threshold is shifted (and
phase separation is generally progressively more suppressed)
the stronger the confinement of the atomic species.

\section{Concluding remarks and future challenges}

In the present work, we have considered the two-component, one-,
two- and three-dimensional nonlinear Schr\"{o}dinger system with
the self-defocusing nonlinearity, and studied the effect of $SO(2)$
rotations on stable stationary dark-bright solitons and higher-dimensional vortex 
complexes. Our numerical findings revealed that the complexes considered 
in this work are robust (over a wide time interval), suggesting possibilities
of observing the underlying states experimentally in Bose-Einstein
condensates. While our starting point was the revisiting of the
simpler (and experimentally observed) dark-dark solitons, we illustrated 
that the transformation and its feature of potentially 
producing breathing states from stationary ones are independent of dimension. 
While analytical solutions are not available in higher dimensions
in order to subject them to the transformation, it is straightforward 
to obtain numerical ones and not only evolve them dynamically, but also
predict on the basis of the difference of their chemical potentials,
the period of the resulting periodic pattern. We performed this step
for a vortex--bright soliton in 2D, obtaining a vortex--vortex state
in that system, while in 3D, the robustness of both vortex-line--soliton
and of vortex-ring--soliton allowed us to form structures with vortex
lines and vortex rings precessing around each other in the two components.

Our observations, while exact in the context of the Manakov model,
as we showcased via select dynamical examples, are no longer
so in the case of unequal interaction coefficients. In fact,
it is evident that in such cases, even weak deviations from the
miscibility-immiscibility threshold that the Manakov system represents
(as is relevant for atomic BECs) may give rise to spontaneously
phase-separating patterns for homogeneous or sufficiently
weakly trapped systems, on top of which the vibration of the
coherent structures of interest may take place. This is a natural
direction for further quantitative exploration, i.e., a more
quantitative identification of the boundaries of robustness
of the states developed herein. Such variation is quite accessible
presently, e.g., via Feshbach resonance techniques \cite{frm}. Additionally,
the realization that the methodology is independent of dimension
and structure also creates the potential of applying features of
this type to other states (including in the focusing case) in order
to obtain other such exotic, time vibrating states. Such studies
are currently in progress and will be reported in the future.

\begin{acknowledgments}
E.G.C. thanks Hans Johnston (UMass) for providing help in connection with
the parallel computing performed in this work. W.W. acknowledges support
from US-NSF (Grant Nos. DMR-1151387). P.G.K gratefully acknowledges support 
from US-NSF under DMS-1312856, and the ERC under FP7, Marie Curie Actions, 
People, International Research Staff Exchange Scheme (IRSES-605096). The 
work of W.W. is supported in part by the Office of the Director of National
Intelligence (ODNI), Intelligence Advanced Research Projects Activity (IARPA),
via MIT Lincoln Laboratory Air Force Contract No.~FA8721-05-C-0002. The 
views and conclusions contained herein are those of the authors and should
not be interpreted as necessarily representing the official policies or 
endorsements, either expressed or implied, of ODNI, IARPA, or the U.S.~Government.
The U.S.~Government is authorized to reproduce and distribute reprints for 
Governmental purpose notwithstanding any copyright annotation thereon. We 
thank the Texas A\&M University for access to their Curie cluster.

\end{acknowledgments}



\begin{thebibliography}{99}
\bibitem{footnote} For instance, Fig.~\ref{fig4d} presents the location of the vortex $(x,y)$
in the first component as a function of time $t$ where the abscissa and
ordinate are denoted by blue and red circles. We compare the numerical
results by using the analytical formulae $x(t)=x_{0}\cos{\left(\omega\,t\right)}$
and $y(t)=y_{0}\sin{\left(\omega\,t\right)}$, with $x_{0}$ and $y_{0}$
standing for the initial location of the complex, while $\omega=\Delta\mu$, as 
shown in section~II. It can be discerned from Fig.~\ref{fig4d} that not only the complex
performs a circular motion, suggesting its periodic nature, but it is
clearly robust.
\end{thebibliography}
\end{document}